\documentclass[11pt, onecolumn]{IEEEtran}

\usepackage{graphicx} 
\usepackage{amsmath} 
\usepackage{amssymb}  
\usepackage{algorithm}
\usepackage{algorithmic}
\usepackage{cite}

\newtheorem{theorem}{Theorem}
\newtheorem{lemma}{Lemma}
\newtheorem{defn}{Definition}
\newtheorem{corollary}{Corollary}
\newtheorem{remark}{Remark}

\IEEEoverridecommandlockouts

\newcommand{\expect}[1]{\mathbb{E}\left[#1\right]}

\newcommand{\prob}[1]{\mathbb{P}\left[#1\right]}

\newcommand{\pth}[1]{\left( #1 \right)}
\newcommand{\qth}[1]{\left[ #1 \right]}
\newcommand{\sth}[1]{\left\{ #1 \right\}}

\newcommand{\indc}[1]{{\mathbf{1}\left\{{#1}\right\}}}

\newcommand{\calE}{{\mathcal{E}}}

\newcommand{\calI}{{\mathcal{I}}}

\newcommand{\calP}{{\mathcal{P}}}

\newcommand{\calR}{{\mathcal{R}}}

\newcommand{\calX}{{\mathcal{X}}}
\newcommand{\calY}{{\mathcal{Y}}}
\newcommand{\calZ}{{\mathcal{Z}}}

\newcommand{\bfF}{{\mathbf{F}}}

\newcommand{\bfR}{{\mathbf{R}}}

\newcommand{\bfX}{{\mathbf{X}}}
\newcommand{\bfY}{{\mathbf{Y}}}
\newcommand{\bfZ}{{\mathbf{Z}}}

\newcommand{\bfx}{{\mathbf{x}}}
\newcommand{\bfy}{{\mathbf{y}}}
\newcommand{\bfz}{{\mathbf{z}}}

\newcommand{\TV}{d_{\mathrm{TV}}}

\renewcommand{\iff}{\Leftrightarrow}

\begin{document}

\title{Universal Joint Image Clustering and Registration using Partition Information \thanks{This work was supported in part by Air Force STTR Contract FA8650-16-M-1819. An initial version of the paper was presented at the 2017 IEEE International Symposium on Information Theory \cite{RamanV2017b}. The authors are with the University of Illinois at Urbana-Champaign.}} 

\author{Ravi Kiran Raman and Lav R.\ Varshney}

\maketitle

\begin{abstract}
We consider the problem of universal joint clustering and registration of images and define algorithms using multivariate information functionals. We first study registering two images using maximum mutual information and prove its asymptotic optimality. We then show the shortcomings of pairwise registration in multi-image registration, and design an asymptotically optimal algorithm based on multiinformation. Further, we define a novel multivariate information functional to perform joint clustering and registration of images, and prove consistency of the algorithm. Finally, we consider registration and clustering of numerous limited-resolution images, defining algorithms that are order-optimal in scaling of number of pixels in each image with the number of images.
\end{abstract}

\section{Introduction}

Suppose you have an unlabeled repository of MRI, CT, and PET scans of brain regions corresponding to different patients from different stages of the diagnostic process. You wish to sort them into clusters corresponding to individual patients and align the images within each cluster. In this work, we address this exact problem of joint clustering and registration of images using novel multivariate information functionals.

Image registration is the task of geometrically aligning two or more images of the same scene taken at different points in time, from different viewpoints, or by different imaging devices. It is a crucial step in most image processing tasks where a comparative study of the different images is required such as medical diagnosis, target detection, image fusion, change detection, and multimodal image restoration. In such applications it is also essential to classify images of different scenes prior to registering images of the same kind. Thus clustering images according to the scene is also critical to computer vision problems such as object tracking, face tagging, cryo-electron microscopy, and remote sensing.

Different digital images of the same scene can appear significantly different from each other, e.g., consider images of a scene that are negatives of each other. Such factors tend to make clustering and registering images harder. Further, such meta-data about the digital images is often not available \emph{a priori}. This emphasizes the need for \emph{universality}, i.e., the design of reliable clustering and registration algorithms that work without the specific knowledge of the priors or channel models that govern the image generation.

Image clustering and registration have often been dealt with separately. However, it is easy to see that clustering registered images and registering images within clusters are both relatively easier. Here we emphasize that the two problems are not separate and define universal, reliable, joint clustering and registration algorithms.

\subsection{Prior Work}

There is rich literature on clustering and registration; we describe a non-exhaustive listing of relevant prior work.

Supervised learning for classification has recently gained prominence through deep convolutional neural networks and other machine learning algorithms. However, these methods require vast amounts of costly labeled training data. Thus, unsupervised image classification is of interest. 

Unsupervised clustering of objects has been studied under numerous optimality and similarity criteria \cite{LuxburgWG2012}. The $k$-means algorithm and its generalization to Bregman divergences \cite{BanerjeeMDG2005} are some popular distance-based methods. Popular techniques for unsupervised image clustering include affinity propagation \cite{DueckF2007}, expectation maximization \cite{KatoZB1999}, independent component analysis \cite{LeeL2002}, and orthogonal subspace projection \cite{RenC2000}. We focus on information-based clustering algorithms \cite{NaganoKI2010, ChanAEKL2015, RamanYV2017}, owing to the ubiquitous nature of information functionals in universal information processing. Universal clustering has been studied in the communication and crowdsourcing \cite{MisraW2013, RamanV2017a}.

Separate from clustering, multi-image registration has been studied extensively \cite{ZitovaF2003}. Prominent region-based registration methods include maximum likelihood (ML) \cite{LeventonG1998}, minimum KL divergence \cite{ChanCYNW2003}, correlation detection \cite{Pratt1974}, and maximum mutual information (MMI) \cite{ViolaW1997, PluimMV2003}. Feature-based techniques have also been explored \cite{AlhichriK2001}.

Lower bounds on mean squared error for image registration in the presence of additive noise using Ziv-Zakai and Cramer-Rao bounds have been explored recently \cite{XuCV2009},\cite{AguerrebereDBS2016}. The MMI decoder was originally developed in universal communication \cite{Goppa1975}. Deterministic reasons for its effectiveness in image registration have been identified \cite{TagareR2015}. Correctness has been established through information-theoretic arguments \cite{ZolleiW2009}. 

A problem closely related to image registration is multireference alignment. There the aim is to denoise a signal from noisy, circularly-translated versions of itself, under Gaussian or binary noise models \cite{BendoryBMZS17_arxiv, AbbePS2017}. Versions of this problem have been considered for image denoising under Gaussian noise \cite{PananjadyWC2017}. Unlike denoising, we consider the task of registration alone but for a wider class of noise models under a universal setting.

\subsection{Our Contributions}

While MMI has been found to perform well in numerous empirical studies, concrete theoretical guarantees are still lacking. In this work, we extend the framework of universal delay estimation \cite{SteinZM1996} to derive universal asymptotic optimality guarantees for MMI in registering two images under the Hamming loss, under mild assumptions on the image models. 

Even though the MMI method is universally asymptotically optimal for registering two images, we show that the method is strictly suboptimal in multi-image registration. We define the max multiinformation (MM) image registration algorithm that uses the multiinformation functional in place of pairwise MMI. We prove that the method is universal and asymptotically optimal using type counting arguments.

Then, we consider the task of joint clustering and registration. We define novel multivariate information functionals to characterize dependence in a collection of images. Under a variety of clustering criteria, we define algorithms using these functionals to perform joint clustering and registration and prove consistency of the methods.

Applications such as cryo-electron microscopy handle a large number of images of several molecular conformations. The task of clustering and registration is critical. With such motivation, we revisit joint clustering and registration under the constraint of limited resolution of images. We define blockwise clustering and registration algorithms, further showing they are order-optimal in the scaling of resolution with number of pixels in the system.

\section{Model}

We now formulate the joint image clustering and registration problem and define the model of images we work with.

\subsection{Image and Noise}

Consider a simple model of images, where each image is a collection of $n$ pixels drawn independently and identically from an unknown prior distribution $P_R(\cdot)$ defined on finite space of pixel values $[r]$. Since the pixels are i.i.d.\ , we represent the original scene of the image by an $n$-dimensional random vector, $\bfR \in [r]^n$. More specifically, the scene is drawn according to $\prob{\bfR} = P_R^{\otimes n}(\bfR)$. 

Consider a finite collection of $\ell$ distinct scenes (drawn i.i.d.\ according to the prior) $\calR = \sth{\bfR^{(1)},\dots,\bfR^{(\ell)}}$. This set can be interpreted as a collection of different scenes. Each image may be viewed as a noisy depiction of an underlying scene.

Consider a collection, $\{\tilde{\bfR}^{(1)},\dots,\tilde{\bfR}^{(m)}\}$, of $m$ scenes drawn from $\calR$, with each scene chosen independently and identically according to the pmf $(p_1,\dots,p_\ell)$, so image $i$ corresponds to a depiction of the scene $\tilde{\bfR}^{(i)}$, for $i \in [m]$.

\begin{figure}[t]
	\centering
	\includegraphics[scale=0.4]{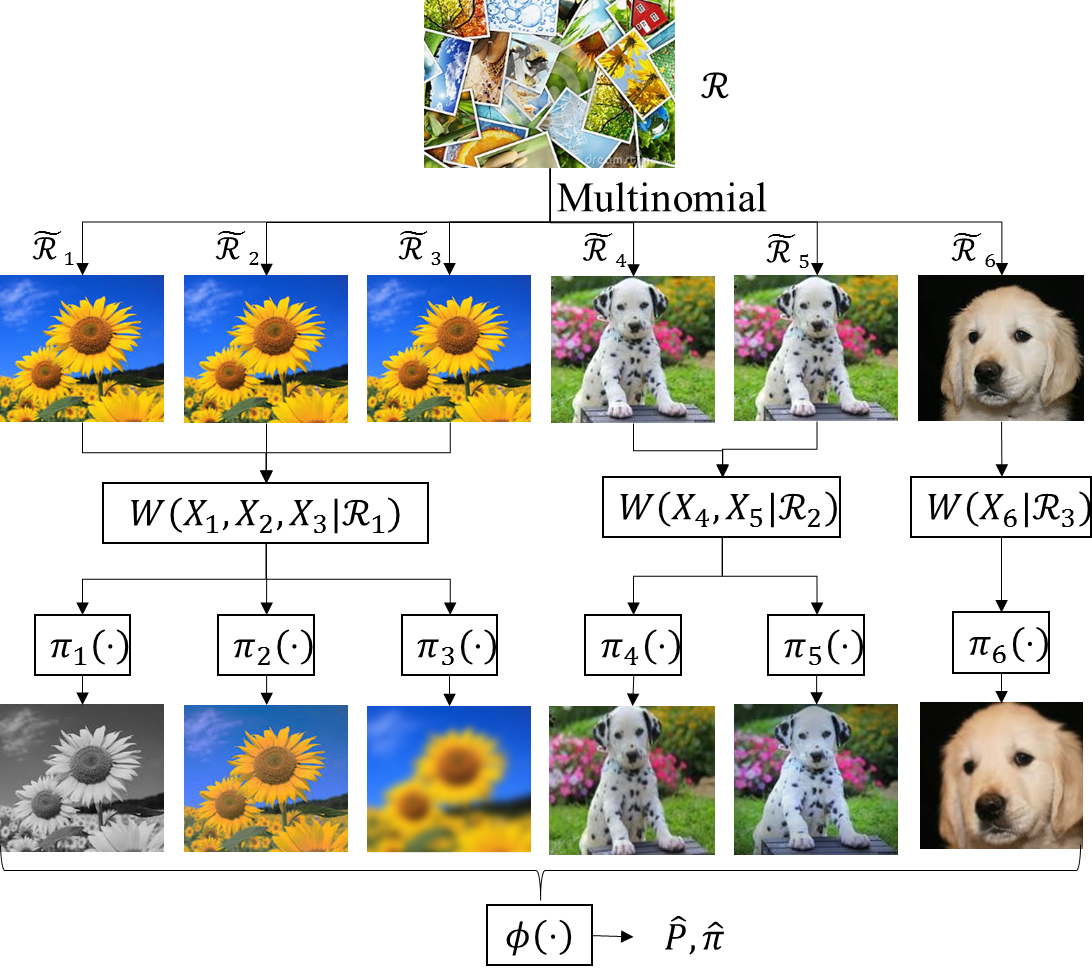}
 	\caption{Model of the joint image clustering and registration problem}
  	\label{fig:image_model}
\end{figure}

We model images corresponding to this collection of scenes as noisy versions of the underlying scenes drawn as follows
\begin{align}
&\prob{\tilde{\bfX}^{[m]} \middle\vert \tilde{\bfR}^{[m]}} = \prod_{j = 1}^\ell \prod_{i = 1}^{n} W\pth{\tilde{X}^{(K(j))}_i \vert R^{(j)}_i}, \label{eqn:image_channel_model}
\end{align}
where $K(j) \subseteq [m]$ is the inclusion-wise maximal subset such that $\tilde{\bfR}^{(i)} = \bfR^{(j)}$ for all $i \in K(j)$. That is, images corresponding to the same scene are jointly corrupted by a discrete memoryless channel, while the images corresponding to different scenes are independent conditioned on the scene. Here we assume $\tilde{\bfX} \in [r]^n$. The system is depicted in Fig. \ref{fig:image_model}.

Consider any two images generated as above. Let $\tilde{W}(y \vert x)$ be the conditional distribution that a pixel in the second image is $y$, given the corresponding pixel in the first image is $x$. Let
\[
\Delta = D(\tilde{W}(Y \vert X_1) \| \tilde{W}(Y \vert X_2) \vert X_1,X_2),
\]
where $X_1,X_2$ are i.i.d.\ copies of pixel values generated according to the marginal distribution of the first image. Note that if $\Delta = \infty$, then with sufficient samples of the image, one can easily identify corresponding pixels in the copy. Hence to avoid the trivial case, we presume there exist finite positive constants $\theta_m, \theta_M$ such that, for any two images,
\begin{equation} \label{eqn:channel_dist_bd}
0 < \theta_m \leq \Delta \leq \theta_M < \infty.
\end{equation}

\subsection{Image Transformations}

Corrupted images are also subject to independent rigid-body transformations such as rotation and translation. Since images are vectors of length $n$, transformations are represented by permutations of $[n]$. Let $\Pi$ be the set of all allowable transformations. We assume $\Pi$ is known.

Let $\pi_j \sim \text{Unif}(\Pi)$ be the transformation of image $j$. Then, the final image is $\bfX^{(j)}_{i} = \tilde{\bfX}^{(j)}_{\pi_j(i)}$, for all $i \in [n]$. Image $\bfX$ transformed by $\pi$ is depicted interchangeably as $\pi(\bfX) = \bfX_{\pi}$.

We assume $\Pi$ forms a \emph{commutative algebra} over the composition operator $\circ$. More specifically,
\begin{enumerate}
\item for $\pi_1,\pi_2 \in \Pi$, $\pi_1 \circ \pi_2 = \pi_2 \circ \pi_1 \in \Pi$;
\item there exists unique $\pi_0 \in \Pi$ s.t. $\pi_0(i) = i$, for all $i \in [n]$;
\item for any $\pi \in \Pi$, there exists a unique inverse $\pi^{-1} \in \Pi$, s.t. $\pi^{-1} \circ \pi = \pi \circ \pi^{-1} = \pi_0$.
\end{enumerate}

The number of distinct rigid transformations of images with $n$ pixels on the $\mathbb{Z}$-lattice is polynomial in $n$, i.e., $|\Pi| = O(n^\alpha)$ for some $\alpha \leq 5$ \cite{NgoKPT2015}.

\begin{defn} \label{defn:permut_cyc}
A \emph{permutation cycle}, $\{i_1,\dots,i_k\}$, is a subset of permutation $\pi$, such that $\pi(i_{j}) = i_{j+1}$, for all $j < k$ and $\pi(i_k) = i_1$.
\end{defn}
It is clear from the pigeonhole principle that any permutation is composed of at least one permutation cycle. Let the number of permutation cycles of a permutation $\pi$ be $\kappa_\pi$.

\begin{defn} \label{defn:iden_block}
\emph{Identity block} of permutation $\pi \in \Pi$ is the inclusion-wise maximal subset $\calI_{\pi} \subseteq [n]$ such that $\pi(i) = i$, for all $i \in \calI_{\pi}$.
\end{defn}
\begin{defn} \label{defn:simple_permut}
A permutation $\pi$ is \emph{simple} if $\kappa_{\pi} = 1$, $\calI_{\pi} = \emptyset$.
\end{defn}
\begin{defn} \label{defn:non_overlap}
Any two permutations $\pi,\pi' \in \Pi$ are said to be \emph{non-overlapping} if $\pi(i) \neq \pi'(i)$ for all $i \in [n]$.
\end{defn}

\begin{lemma} \label{lemma:small_iden_block}
Let $\pi$ be chosen uniformly at random from the set of all permutations of $[n]$. Then for any constants $c \in (0,1],C$, 
\begin{equation}
\prob{|\calI_{\pi}| > cn} \lesssim \exp\pth{-cn}, \quad \prob{\kappa_{\pi} > C\tfrac{n}{\log n}} = o(1).
\end{equation}
\end{lemma}
\begin{IEEEproof}
First, we observe that the number of permutations that have an identity block of size at least $cn$ is given by
\begin{align*}
\nu_c &\leq \binom{n}{cn}((1-c)n)! = \frac{n!}{(cn)!}.
\end{align*}
Thus, from Stirling's approximation,
\begin{align*}
\prob{|\calI_{\pi}| \geq cn} &\leq \tfrac{1}{\sqrt{2\pi}}\exp\pth{-(cn+\tfrac{1}{2})\log(cn) + cn}.
\end{align*}

Lengths and number of cycles in a random permutation may be analyzed as detailed in \cite{SheppL1966}. In particular, we note that for a random permutation $\pi$, $\expect{\kappa_{\pi}} = \log n + O(1)$. Using Markov's inequality, the result follows.
\end{IEEEproof}

Following Lem.~\ref{lemma:small_iden_block}, we assume that for any $\pi \in \Pi$, $\kappa_{\pi} = o\pth{n/\log(n)}$, i.e., the number of permutation cycles does not grow very fast. Further, let $|\calI_{\pi}| = o(n)$ for any $\pi \in \Pi$.

\subsection{Performance Metrics}

We now introduce formal metrics to quantify performance of the joint clustering and registration algorithms. 

\begin{defn} \label{defn:opt_clust}
A \emph{clustering} of images $\{\bfX^{(1)},\dots,\bfX^{(m)}\}$ is a partition $P$ of $[m]$. The sets of a partition are referred to as \emph{clusters}. The clustering is said to be \emph{correct} if
\[
i,j \in C \iff \bfR^{(i)} = \bfR^{(j)}, \text{ for all } i,j \in [m],~ C \in P. 
\]
\end{defn}
Let $\calP$ be the set of all partitions of $[m]$. For a given collection of scenes, we represent the correct clustering by $P^{*}$.

\begin{defn}
A partition $P$ is \emph{finer} than $P'$, $P \preceq P'$, if $\text{ for all } C\in P, \text{ there exists } C'\in P' : C \subseteq C'$. Similarly, a partition $P$ is \emph{denser} than $P'$, $P \succeq P'$, if $P' \preceq P$.
\end{defn}

\begin{defn} \label{defn:opt_regn}
The \emph{correct registration} of an image $\bfX$ transformed by $\pi \in \Pi$ is $\hat{\pi} = \pi^{-1}$.
\end{defn}

\begin{defn}
A \emph{universal clustering and registration algorithm} is a sequence of functions $\Phi^{(n)}: \sth{\bfX^{(1)},\dots,\bfX^{(m)}} \rightarrow \calP \times \Pi^m$ that are designed in the absence of knowledge of $W$, $\{p_1,\dots,p_\ell\}$, and $P_R$. Here the index $n$ corresponds to the number of pixels in each image.	
\end{defn}

We focus on the $0$-$1$ loss function to quantify performance. 
\begin{defn}
The \emph{error probability} of an algorithm $\Phi^{(n)}$ that outputs $\hat{P} \in \calP$, $\pth{\hat{\pi}_1,\dots,\hat{\pi}_m} \in \Pi^m$ is
\begin{align}
P_e(\Phi^{(n)}) &= \prob{\sth{\cup_{i \in [m]} \{\hat{\pi}_i \neq \pi_i^{-1}\}} \cup \{\hat{P} \neq P^*\}}. \label{eqn:clustering_error}
\end{align} 
\end{defn}

\begin{defn}
Alg.~$\Phi^{(n)}$ is \emph{asymptotically consistent} if $\lim_{n \rightarrow \infty} P_e(\Phi^{(n)}) = 0$, and is \emph{exponentially consistent} if $\lim_{n \rightarrow \infty} -\log P_e(\Phi^{(n)}) > 0$.
\end{defn}

\begin{defn}
The \emph{error exponent} of an algorithm $\Phi^{(n)}$ is
\begin{equation}
\calE(\Phi^{(n)}) = \lim_{n \rightarrow \infty} -\tfrac{1}{n} \log P_e(\Phi^{(n)}).
\end{equation}
\end{defn}
We use $\Phi$ to denote $\Phi^{(n)}$ when clear from context. 

\section{Registration of two images}

\begin{figure}[t]
	\centering
	\includegraphics[scale=0.7]{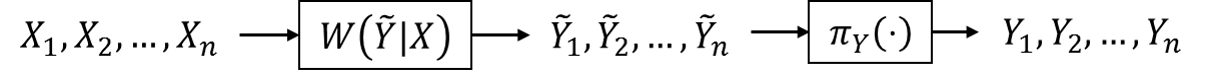}
 	\caption{Model of the two image registration problem: Image $\bfY$ is to be registered with source image $\bfX$.}
  	\label{fig:two_image_regn}
\end{figure}

We first consider the problem of registering two images, i.e., $m = 2$, $\ell = 1$. Thus the problem reduces to registering an image $\bfY$ obtained as a result of transforming the output of an equivalent discrete memoryless channel $W$, given input image (reference) $\bfX$. The channel model is depicted in Fig.~\ref{fig:two_image_regn}.

This problem has been well-studied in practice and a heuristic method used is the MMI method defined as
\begin{equation} \label{eqn:MMI_regn}
\hat{\pi}_{\text{MMI}} = \arg\max_{\pi\in\Pi} \hat{I}(X;Y_{\pi}),
\end{equation}
where $\hat{I}(X;Y)$ is the mutual information of the empirical distribution $\hat{P}(x,y) = \tfrac{1}{n} \sum_{i=1}^n \indc{X_i = x, Y_i = y},$ where $\indc{\cdot}$ is the indicator function. Note that the MMI method is universal.

Since transformations are chosen uniformly at random, the maximum likelihood (ML) estimate is Bayes optimal:
\begin{equation}\label{eqn:ML_regn}
\hat{\pi}_{\text{ML}} = \arg\max_{\pi\in\Pi} \prod_{i=1}^n W\pth{Y_{\pi(i)} \vert X_i}.
\end{equation}
We first show that the MMI method, and consequently the ML method, are exponentially consistent. We then show that the error exponent of MMI matches that of ML.

\subsection{Exponential Consistency}

The empirical mutual information of i.i.d.\ samples is asymptotically exponentially consistent.
\begin{theorem} \label{thm:ML_info_conv}
Let $\sth{(X_1,Y_1),\dots,(X_n,Y_n)}$ be random samples drawn i.i.d.\ according to the joint distribution $p$ defined on the finite space $\calX\times\calY$. For fixed alphabet sizes, $|\calX|,|\calY|$, the ML estimate of entropy and mutual information are asymptotically exponentially consistent and satisfy
\begin{equation}
\prob{|\hat{H}(X) - H(X)| > \epsilon} \leq (n+1)^{|\calX|} e^{-c n \epsilon^4},
\end{equation}
\begin{equation}
\prob{|\hat{I}(X;Y) - I(X;Y)| > \epsilon} \leq 3(n+1)^{|\calX||\calY|} e^{-\tilde{c} n \epsilon^4}, \label{eqn:inf_cont}
\end{equation}
where $c = \pth{2|\calX|^2\log2}^{-1}$, $\tilde{c} = \pth{32\max\{|\calX|,|\calY|\}^2\log2}^{-1}$.
\end{theorem}
\begin{IEEEproof}
The proof is given in \cite[Lem.~7]{RamanV2016_arxiv}.
\end{IEEEproof}

\begin{theorem} \label{thm:MMI_ML_consistency}
MMI and ML are exponentially consistent.
\end{theorem}
\begin{IEEEproof}
Let $\Phi_{\text{MMI}}(\bfX,\bfY) = \hat{\pi}_{\text{MMI}}$ and let the correct registration be $\pi^*$. Then,
\begin{align}
&P_e(\Phi_{\text{MMI}}) = \prob{\hat{\pi}_{\text{MMI}} \neq \pi^*} \notag \\
&\quad \leq \sum_{\pi \in \Pi} \prob{\hat{I}(X;Y_\pi)>\hat{I}(X;\tilde{Y})} \label{eqn:thm2_union_bd} \\
&\quad \leq \sum_{\pi \in \Pi} \prob{|\hat{I}(X;Y_\pi) - (\hat{I}(X;\tilde{Y}) - I(X;\tilde{Y}))| > I(X;\tilde{Y})} \notag \\
&\quad \leq \sum_{\pi \in \Pi} \prob{\hat{I}(X;Y_\pi) + |\hat{I}(X;\tilde{Y}) - I(X;\tilde{Y})| > I(X;\tilde{Y})} \label{eqn:thm2_trng_ineq} \\
&\quad \leq 2|\Pi|\exp\sth{-Cn(I(X;\tilde{Y}))^4}, \label{eqn:thm2_info_conv}
\end{align}
where (\ref{eqn:thm2_union_bd}), (\ref{eqn:thm2_trng_ineq}), and (\ref{eqn:thm2_info_conv}) follow from the union bound, the triangle inequality, and (\ref{eqn:inf_cont}), respectively. Here $C \asymp r^{-2}$ is a constant.

Thus MMI is exponentially consistent as $|\Pi| = O(n^\alpha)$. Finally, $P_e(\Phi_{\text{ML}}) \leq P_e(\Phi_{\text{MMI}})$ and thus, the ML estimate is also exponentially consistent.
\end{IEEEproof}

Thm.~\ref{thm:MMI_ML_consistency} implies there exists $\epsilon > 0$ such that 
\[
\calE(\Phi_{\text{ML}}) \geq \calE(\Phi_{\text{MMI}}) \geq \epsilon.
\]

\subsection{Whittle's Law and Markov Types}

We now summarize a few results on the number of types and Markov types which are eventually used to analyze the error exponent of image registration.

Consider a sequence $\bfx \in \calX^n$. The empirical distribution $q_X$ of $\bfx$ is the \emph{type} of the sequence. Let $X \sim q_X$ be a dummy random variable. Let $T_X^n$ be the set of all sequences of length $n$, of type $q_X$. The number of possible types of sequences of length $n$ is polynomial in $n$, i.e., $O(n^{|\calX|})$ \cite{Csiszar1998}. 

The number of sequences of length $n$, of type $q_X$, is
\[
|T_X^n| = \frac{n!}{\prod_{a \in \calX} (nq_X(a))!}.
\] 
From bounds on multinomial coefficients, the number of sequences of length $n$ and type $q$ is bounded as \cite{Csiszar1998}
\begin{equation} \label{eqn:type0_bound}
(n+1)^{-|\calX|}2^{nH(X)} \leq |T_X^n| \leq 2^{nH(X)}.
\end{equation}

Consider a Markov chain defined on the space $[k]$. Given a sequence of $n+1$ samples from $[k]$ we can compute the matrix $\bfF$ of transition counts, where $F_{ij}$ corresponds to the number of transitions from state $i$ to state $j$. By Whittle's formula \cite{Whittle1955}, the number of sequences $(a_1,\dots,a_{n+1})$ with $a_i \in [k], i \in [n+1]$, with $a_1 = u$ and $a_{n+1} = v$ is
\begin{equation}\label{eqn:Whittle}
N_{uv}^{(n)}(F) = \prod_{i\in[k]}\frac{(\sum_{j\in[k]}F_{ij})!}{\prod_{j\in[k]}F_{ij}!} G_{vu}^*,
\end{equation}
where $G_{vu}^*$ corresponds to the $(v,u)$th cofactor of the matrix $G = \{g_{ij}\}_{i,j\in[k]}$ with
\[
g_{ij} = \indc{i = j} - \frac{F_{ij}}{\sum_{j\in[k]} F_ij}.
\]

The first-order Markov type of a sequence $\bfx \in \calX^n$ is defined as the empirical distribution $q_{X_0,X_1}$, given by
\[
q_{X_0,X_1}(a_0,a_1) = \frac{1}{n} \sum_{i=1}^n \indc{(x_i,x_{i+1}) = (a_0,a_1)}.
\]
Here we assume that the sequence is cyclic with period $n$, i.e., for any $i > 0, x_{n+i} = x_{i}$. Let $(X_0,X_1) \sim q_{X_0,X_1}$. Then, from (\ref{eqn:Whittle}), the set of sequences of type $q_{X_0,X_1}$, $T_{X_0,X_1}^n$, satisfies
\[
|T_{X_0,X_1}^n| = \pth{\sum_{a \in \calX} G_{a,a}^*} \prod_{a_0 \in \calX}\frac{(nq_{0}(a_0))!}{\prod_{a_1 \in \calX}(nq_{\bfx_0,\bfx_1}(a_0,a_1))!} .
\]
From the definition of $G$, we can bound the trace of the cofactor matrix of $G$ as
\[
\frac{|\calX|}{(n+1)^{|\calX|}} \leq \sum_{a \in \calX} G_{a,a}^* \leq |\calX|.
\]
Again using the bounds on multinomial coefficients, we have
\begin{align}
&|\calX|(n+1)^{-(|\calX|^2+|\calX|)} 2^{n(H(X_0,X_1) - H(X_0))} \notag\\
&\qquad \leq |T_{X_0,X_1}^n| \leq |\calX| 2^{n(H(X_0,X_1) - H(X_0))}. \label{eqn:type1_bound}
\end{align}

The joint first-order Markov type of a pair of sequences $\bfx \in \calX^n$, $\bfy \in \calY^n$ is the empirical distribution
\[
q_{X_0,X_1,Y}(a_0,a_1,b) = \frac{1}{n} \sum_{i=1}^n \indc{(x_i,x_{i+1},y_i) = (a_0,a_1,b)}.
\]
Then given $\bfx$, the set of conditional first-order Markov type sequences, $T_{Y \vert X_0,X_1}^n(\bfx)$ satisfies \cite{SteinZM1996}
\begin{align} 
&(n+1)^{-|\calX|^2|\calY|} 2^{n(H(X_0,X_1,Y) - H(X_0,X_1))} \notag \\
& \qquad \leq |T_{Y \vert X_0,X_1}^n(\bfx)| \leq 2^{n(H(X_0,X_1,Y) - H(X_0,X_1))}. \label{eqn:cond_Markov_type}
\end{align}
 
\begin{lemma} \label{lemma:perut_Markov_map}
Let $\pi_1,\pi_2 \in \Pi$ be any two non-overlapping permutations and let $\bfx_{\pi_1}, \bfx_{\pi_2}$ be the corresponding permutations of $\bfx$. Let $\pi_1^{-1}\circ \pi_2 \in \Pi$ be a simple permutation. Then, for every $\bfx$, there exists $\tilde{\bfx}$, such that
\begin{equation}
|T_{X_{\pi_1},X_{\pi_2}}^n| = |T_{X_0,X_1}^n|, \quad |T_{Y \vert X_{\pi_1},X_{\pi_2}}^n(\bfx)| = |T_{Y \vert X_0,X_1}^n(\tilde{\bfx})|. \notag
\end{equation}
\end{lemma}
\begin{IEEEproof}
Since permutations are non-overlapping, there is a bijection from $T_{X_{\pi_1},X_{\pi_2}}^n$ to $T_{X_0,X_1}^n$, where $(X_0,X_1) \sim q_{X_{\pi_1},X_{\pi_2}}$. Specifically, consider the permutation $\pi \in \Pi$ defined iteratively as $\pi(i+1) = \pi_2(\pi_1^{-1}(\pi(i)))$, with $\pi(1) = \pi_1(1)$. Then, for any $\bfx \in T_{X_{\pi_1},X_{\pi_2}}^n$, the sequence $\bfx_{\pi} \in T_{X_0,X_1}^n$. Further, this map is invertible and so the sets are of equal size. 

Result for conditional types follows \emph{mutatis mutandis}.
\end{IEEEproof}

Lem.~\ref{lemma:perut_Markov_map} implies $|T_{X_{\pi_1},X_{\pi_2}}^n|$ and $|T_{Y \vert X_{\pi_1},X_{\pi_2}}^n(\bfx)|$ satisfy (\ref{eqn:type1_bound}) and (\ref{eqn:cond_Markov_type}) respectively. We now show that the result of Lem.~\ref{lemma:perut_Markov_map} can be extended to any two permutations $\pi_1,\pi_2 \in \Pi$.
\begin{lemma} \label{lemma:permut_generic_count}
Let $\pi_1,\pi_2 \in \Pi$. For any $\bfx$,
\[
|T_{X_{\pi_1},X_{\pi_2}}^n| = 2^{n \pth{H(q_{X_{\pi_1},X_{\pi_2}}) - H(q_{X}) + o(1)}}.
\]
\end{lemma}
\begin{IEEEproof}
Let $\pi = \pi_1^{-1} \circ \pi_2$ and $\kappa = \kappa_{\pi}$. For $i \in [\kappa]$ let the length of permutation cycle $i$ of $\pi$ be $\alpha_i n$ for $\alpha_i \in (0,1]$. Further, $\sum_{i=1}^{\kappa} \alpha_i \leq 1$. Let $\calI_{\pi}$ be the identity block of $\pi$ and let $\gamma = \gamma_{\pi}$. Then we have the decomposition
\begin{equation} \label{eqn:lemma_2_decomp}
q_{X_{\pi_1},X_{\pi_2}}(a_0,a_1) = \sum_{i=1}^\kappa \alpha_i q_i(a_0,a_1) + \gamma q_{\calI}(a_0,a_1),
\end{equation}
for all $(a_0,a_1) \in \calX^2$. Here, $q_i$ is the first-order Markov type defined on the $i$th permutation cycle of $\pi$ and $q_{\calI}$ is the zeroth-order Markov type corresponding to the identity block of $\pi$.

From (\ref{eqn:lemma_2_decomp}), we see that given a valid decomposition of types $\{q_i\}, q_{\calI}$, the number of sequences can be computed as a product of the number of subsequences of each type, i.e.,
\[
|T_{X_{\pi_1},X_{\pi_2}}^n| = \sum |T_{q_{\calI}}^{\gamma n}| \prod_{i=1}^{\kappa} |T_{q_i}^{\alpha_i n}|,
\]
where the sum is over all valid decompositions in (\ref{eqn:lemma_2_decomp}).

Additionally, from Lem.~\ref{lemma:perut_Markov_map} we know the number of valid subsequences of each type. Let $q_i'$ be the marginal corresponding to the first-order Markov type $q_i$.

Thus, we upper bound the size of the set as
\begin{align}
&|T_{X_{\pi_1},X_{\pi_2}}^n| \leq \sum 2^{\gamma n H(q_{\calI})} \prod_{i=1}^{\kappa} |\calX| 2^{\alpha_i n \pth{H(q_i) - H(q_i')}} \label{eqn:lemma_2_ub_ref} \\
& \quad \leq |\calX|^{\kappa} (\gamma n + 1)^{|\calX|} \prod_{i=1}^{\kappa} (\alpha_i n + 1)^{|\calX|^2} 2^{n M(q_{X_{\pi_1},X_{\pi_2}}) } \label{eqn:lemma_2_poly_type},
\end{align}
where 
\[
M(q_{X_{\pi_1},X_{\pi_2}}) = \max \gamma H(q_{\calI}) + \sum_{i=1}^{\kappa} \alpha_i (H(q_i) - H(q_i')),
\]
the maximum taken over all valid decompositions in (\ref{eqn:lemma_2_decomp}). Here, (\ref{eqn:lemma_2_ub_ref}) follows from (\ref{eqn:type0_bound}) and (\ref{eqn:type1_bound}), and (\ref{eqn:lemma_2_poly_type}) follows since the total number of possible types is polynomial in the length of the sequence. Since $\kappa = o\pth{\tfrac{n}{\log(n)}}$,
\[
|T_{X_{\pi_1},X_{\pi_2}}^n| \leq 2^{n \pth{M(q_{X_{\pi_1},X_{\pi_2}}) + o(1)}}.
\]

Now, let $q_i''(a_0,a_1) = \frac{1}{|\calX|}q_i'(a_1)$, for all $i \in [\kappa]$, and $(a_0,a_1) \in \calX^2$. And let $q = q_{X_{\pi_1},X_{\pi_2}}$. Since $\gamma = o(1)$, using Jensen's inequality, we have
\begin{align*}
\sum_{i=1}^{\kappa} \alpha_i (H(q_i) - H(q_i')) &= \sum_{i=1}^{\kappa} \alpha_i \qth{\log (|\calX|) - D\pth{q_i \| q_i''}} \\
& \leq \log (|\calX|) - D\pth{q\|q''} \\
& = H(q) - H(q').
\end{align*}
Thus,
\[
|T_{X_{\pi_1},X_{\pi_2}}^n| \leq 2^{n \pth{H(q) - H(q') + o(1)}}.
\]

To obtain the lower bound, we note that the total number of sequences is at least the number of sequences obtained from any single valid decomposition. Thus from (\ref{eqn:type0_bound}) and (\ref{eqn:type1_bound}),
\[
|T_{X_{\pi_1},X_{\pi_2}}^n| \geq 2^{n \pth{M(q_{X_{\pi_1},X_{\pi_2}}) + o(1)}}. 
\]

Now, for large $n$, consider $S = \{i \in \kappa: \alpha_i n = \Omega(n^\beta)\}$ for some $\beta > 0$. Any other cycle of smaller length contributes $o(1)$ to the exponent due to Lem.~\ref{lemma:small_iden_block}. One valid decomposition of (\ref{eqn:lemma_2_decomp}) is to have $q_i = q$, for all $i \in [\kappa]$. However, the lengths of the subsequences are different and $q$ may not be a valid type of the corresponding length. Nevertheless, for each $i \in S$, there exists a type $q_i$ such that $\TV(q_i,q) \leq \tfrac{|\calX|}{2\alpha_i n}$, where $\TV(\cdot)$ is the total variational distance. Further, entropy is continuous in the distribution \cite{NetrapalliBSS2010} and satisfies
\[
|H(q_i) - H(q)| \leq \frac{|\calX|}{\alpha_i n} \log \pth{\alpha_i n} = o(1).
\]
This in turn indicates that
\[
|T_{X_{\pi_1},X_{\pi_2}}^n| \geq 2^{n \pth{H(q) - H(q') + o(1)}}.
\]
\end{IEEEproof}
Similar decomposition follows for conditional types as well.

\subsection{Error Analysis}

We are interested in the error exponent of MMI-based image registration, in comparison to ML. We first note that the error exponent of the problem is characterized by the pair of transformations that are the hardest to compare.

Define $\Psi_{\pi,\pi'}$ as the binary hypothesis testing problem corresponding to image registration when the allowed transformations are only $\{\pi,\pi'\}$. Let $P_{\pi,\pi'}(\Phi), \calE_{\pi,\pi'}(\Phi)$ be the corresponding error probability and error exponent.
\begin{lemma} \label{lemma:multi_to_binary}
Let $\Phi$ be an asymptotically exponentially consistent estimator. Then,
\begin{equation}\label{eqn:multi_to_binary}
\calE(\Phi) = \min_{\pi,\pi'\in\Pi} \calE_{\pi,\pi'}(\Phi).
\end{equation}
\end{lemma}
\begin{IEEEproof}
Let $\hat{\pi}$ be the estimate output by $\Phi$ and $\pi^* \sim \text{Unif}(\Pi)$ be the correct registration. We first upper bound the error probability as
\begin{align}
P_e(\Phi) &= \prob{\hat{\pi} \neq \pi^*} \leq \sum_{\pi \neq \pi'} \tfrac{1}{|\Pi|} \prob{\hat{\pi} = \pi \vert \pi^* = \pi'} \label{eqn:lemma2_union_bd} \\
&\leq \frac{2}{|\Pi|}\sum_{\pi \neq \pi'} P_{\pi,\pi'}(\Phi), \notag
\end{align}
where (\ref{eqn:lemma2_union_bd}) follows from the union bound.

Additionally, we have
\begin{align*}
P_e(\Phi) &= \frac{1}{|\Pi|}\sum_{\pi \in \Pi} \sum_{\pi' \in \Pi\backslash\{\pi\}} \prob{\hat{\pi} = \pi' \vert \pi^* = \pi} \\
&\geq \frac{1}{|\Pi|}\max_{\pi,\pi' \in \Pi} P_{\pi,\pi'}(\Phi).
\end{align*}

Finally, since $|\Pi| = O(n^\alpha)$, the result follows.
\end{IEEEproof}

Thus, it suffices to consider the binary hypothesis tests to study the error exponent of image registration.
\begin{theorem}\label{thm:error_conv_binary}
Let $\{\pi_1,\pi_2\} \subseteq \Pi$. Then,
\begin{equation}\label{eqn:error_conv_binary}
\lim_{n \rightarrow \infty} \frac{P_{\pi_1,\pi_2}(\Phi_{\text{MMI}})}{P_{\pi_1,\pi_2}(\Phi_{\text{ML}})} = 1.
\end{equation}
\end{theorem}
\begin{IEEEproof}
Probabilities of i.i.d.\ sequences are defined by their joint type. Thus, we have
\begin{align}
P_{\pi_1,\pi_2}(\Phi_{\text{MMI}}) &= \prob{\hat{\pi}_{\text{MMI}} \neq \pi^*} \notag\\
&= \sum_{\bfx \in \calX^n}\sum_{\bfy \in \calY^n} \prob{\bfx,\bfy}\indc{\hat{\pi}_{\text{MMI}} \neq \pi^*} \notag\\
&= \sum_{q} \pth{\prod_{a \in \calX}\prod_{b \in \calY} (\prob{a,b})^{nq(a,b)}} \nu_{\text{MMI}}(q), \label{eqn:thm3_typical_seq}
\end{align}
where the summation in (\ref{eqn:thm3_typical_seq}) is over the set of all joint types of sequences of length $n$ and $\nu_{\text{MMI}}(q)$ is the number of sequences $(\bfx,\bfy)$ of length $n$ with joint type $q$ such that the MMI algorithm makes a decision error. 

If a sequence $\bfy \in T_{Y \vert X_{\pi_1},X_{\pi_2}}^n(\bfx)$ is in error, then all sequences in $T_{Y \vert X_{\pi_1},X_{\pi_2}}^n(\bfx)$ are in error as $\bfx$ is drawn from an i.i.d.\ source. Now, we can decompose the number of sequences under error as follows
\begin{align*}
\nu(q) &= \sum_{\bfx \in \calX^n} \sum_{T_{Y \vert X_{\pi_1},X_{\pi_2}}^n \subseteq T_{Y \vert X}^n : \text{error}} |T_{Y \vert X_{\pi_1},X_{\pi_2}}^n(\bfx)| \\
&= \sum_{T_{X_{\pi_1},X_{\pi_2}}^n \subseteq T_X^n} |T_{X_{\pi_1},X_{\pi_2}}| \sum |T_{Y \vert X_{\pi_1},X_{\pi_2}}^n|,
\end{align*}
where the sum is taken over $T_{Y \vert X_{\pi_1},X_{\pi_2}}^n \subseteq T_{Y \vert X}^n$ such that there is a decision error. The final line follows from the fact that given the joint first-order Markov type, the size of the conditional type is independent of the exact sequence $\bfx$.

Finally, we have
\begin{align}
P_{\pi_1,\pi_2}(\Phi_{\text{MMI}}) &= \sum_{q} \prod_{a \in \calX}\prod_{b \in \calY} (\prob{a,b})^{nq(a,b)} \nu_{\text{ML}}(q) \qth{\frac{\nu_{\text{MMI}}(q)}{\nu_{\text{ML}}(q)}} \notag\\
&\leq P_{\pi_1,\pi_2}(\Phi_{\text{ML}}) \max_{q}\sth{\frac{\nu_{\text{MMI}}(q)}{\nu_{\text{ML}}(q)}}.
\end{align}

The result then follows from the forthcoming Lem.~\ref{lemma:no_ratio}.
\end{IEEEproof}

\begin{lemma} \label{lemma:no_ratio}
\begin{equation}
\lim_{n \rightarrow \infty} \max_{q}\sth{\frac{\nu_{\text{MMI}}(q)}{\nu_{\text{ML}}(q)}} = 1.
\end{equation}
\end{lemma}
\begin{IEEEproof}
Observe that for images with i.i.d.\ pixels, MMI is the same as minimizing the joint entropy, i.e.,
\begin{align*}
\max_{\pi \in \Pi} \hat{I}(X;\pi(Y)) &= \max_{\pi \in \Pi} \hat{H}(X) + \hat{H}(\pi(Y)) - \hat{H}(X,\pi(Y)) \\
&= \hat{H}(X) + \hat{H}(Y) - \min_{\pi \in \Pi} \hat{H}(X,\pi(Y)).
\end{align*}

Further we know that there is a bijective mapping between sequences corresponding to the permutations and the sequences of the corresponding first-order Markov type from Lem.~\ref{lemma:perut_Markov_map} and \ref{lemma:permut_generic_count}. Thus, the result follows from \cite[Lem.~1]{SteinZM1996}.
\end{IEEEproof}

\begin{theorem}\label{thm:error_exponent_two_image}
\begin{equation}
\calE(\Phi_{\text{MMI}}) = \calE(\Phi_{\text{ML}}).
\end{equation}
\end{theorem}
\begin{IEEEproof}
This follows from Thm.~\ref{thm:error_conv_binary} and Lem.~\ref{lemma:multi_to_binary}.
\end{IEEEproof}

Thus, we can see that using MMI for image registration is not only universal, but also asymptotically optimal. We next study the problem of image registration for multiple images.

\section{Multi-image Registration}

\begin{figure}[t]
	\centering
	\includegraphics[scale=0.7]{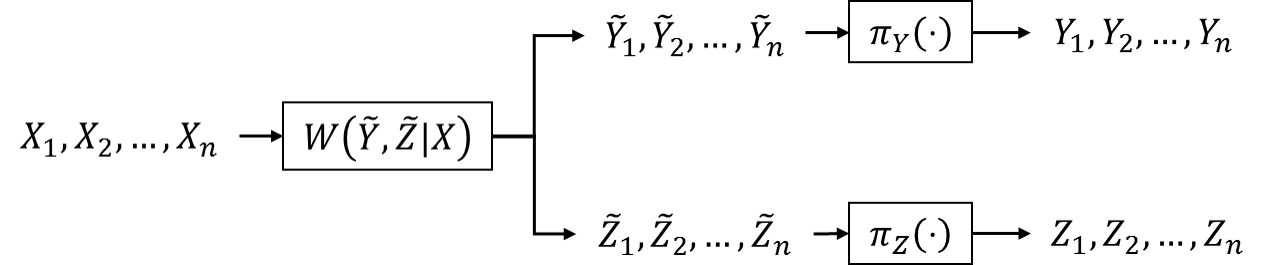}
 	\caption{Model of the three-image registration problem: Images $\bfY, \bfZ$ are to be registered with source image $\bfX$.}
  	\label{fig:three_image_regn}
\end{figure}

Having universally registered two images, we now consider aligning multiple copies of the same image. For simplicity, let us consider aligning three images; ; results can directly be extended to any finite number of images. Let $\bfX$ be the source image and $\bfY,\bfZ$ be the noisy, transformed versions to be aligned as shown in Fig.~\ref{fig:three_image_regn}.

Here, the ML estimates are
\begin{equation}\label{eqn:three_image_ML}
(\hat{\pi}_{Y,\text{ML}},\hat{\pi}_{Z,\text{ML}}) = \arg\max_{\pi_1,\pi_2} \prod_{i=1}^n W\pth{Y_{\pi_1(i)},Z_{\pi_2(i)} \vert X_i}.
\end{equation}

\subsection{MMI is not optimal}

We know MMI is asymptotically optimal at aligning two images. Is pairwise MMI, i.e.
\begin{equation}\label{eqn:pairwise_MMI}
\hat{\pi}_Y = \arg\max_{\pi \in \Pi} \hat{I}(X;Y_{\pi}), ~ \hat{\pi}_Z = \arg\max_{\pi \in \Pi} \hat{I}(X;Z_{\pi}),
\end{equation}
optimal for multi-image registration? We show pairwise MMI is suboptimal even though individual transformations are chosen independently and uniformly from $\Pi$.

\begin{theorem}\label{thm:pairwise_sub_optimal}
There exists channel $W$ and prior, such that pairwise MMI is suboptimal for multi-image registration.
\end{theorem}
\begin{IEEEproof}
Let $X_i \stackrel{\text{i.i.d.}}{\sim} \text{Bern}(1/2), i \in [n]$. Consider physically degraded images $\bfY,\bfZ$ obtained as outputs of the channel $ W\pth{y,z \vert x} = W_1\pth{y \vert x} W_2\pth{z \vert y}$. This is depicted in Fig. \ref{fig:phy_degraded}.

\begin{figure}[b]
	\centering
	\includegraphics[scale=0.7]{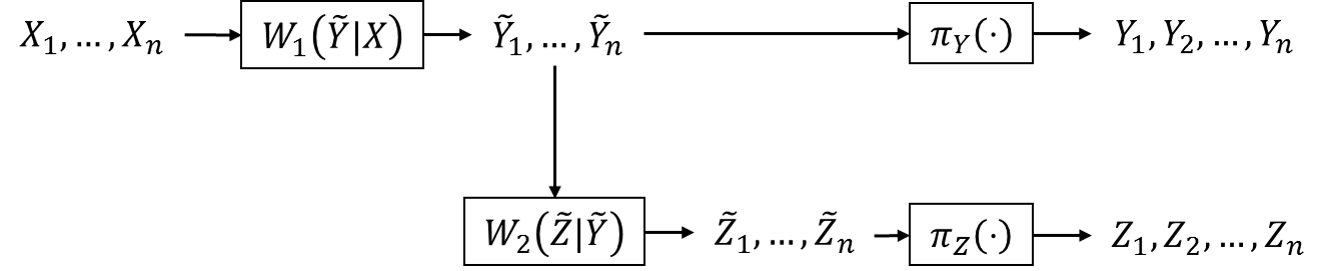}
 	\caption{Three-image registration problem with image $\bfZ$ a physically-degraded version of $\bfY$ and to be registered with source image $\bfX$.}
  	\label{fig:phy_degraded}
\end{figure}

Naturally, the ML estimate of the transformations is obtained by registering image $\bfZ$ to $\bfY$ and subsequently registering $\bfY$ to $\bfX$, instead of registering each of the images pairwise to $\bfX$. That is, from (\ref{eqn:three_image_ML})
\begin{align*}
&\pth{\hat{\pi}_{Y,\text{ML}}, \hat{\pi}_{Z,\text{ML}}} \\
&\quad = \arg\max_{(\pi_1,\pi_2) \in \Pi^2} \prod_{i=1}^n W_1(Y_{\pi_1(i)} \vert X_i)W_2(Z_{\pi_2(i)} \vert Y_{\pi_1(i)}).
\end{align*}
It is evident that $\pi_Z$ is estimated based on the estimate of $\pi_Y$. 

Let $\calE_W(\Phi_{\text{ML}})$ be the error exponent of $\Phi_{\text{ML}}$ for the physically-degraded channel. Since the ML estimate involves registration of $\bfY$ to $\bfX$ and $\bfZ$ to $\bfY$, the error exponent is $\calE_W(\Phi_{\text{ML}}) = \min\sth{\calE_{W_1}(\Phi_{\text{ML}}), \calE_{W_2}(\Phi_{\text{ML}})}$.

Let $\calE_{Q}(\Phi_{\text{MMI}})$ be the error exponent of MMI. Then, error exponent of pairwise MMI is $\calE(\Phi_{\text{MMI}}) = \min \sth{\calE_{W_1}(\Phi_{\text{MMI}}), \calE_{W_1*W_2}(\Phi_{\text{MMI}})}$. We know MMI is asymptotically optimal for two image registration and so $\calE_{W_1}(\Phi_{\text{MMI}}) = \calE_{W_1}(\Phi_{\text{ML}}), ~ \calE_{W_1*W_2}(\Phi_{\text{MMI}}) = \calE_{W_1*W_2}(\Phi_{\text{ML}})$.

More specifically, let $W_1 = BSC(\alpha)$ and $W_2 = BSC(\beta)$ for some $\alpha,\beta \in (0,1/2)$. Let, $W_1*W_2 = BSC(\gamma)$, where $\gamma = \alpha(1-\beta)+(1-\alpha)\beta > \max\sth{\alpha,\beta}$. Then,
\begin{align}
\calE(\Phi_{\text{MMI}}) &\leq \calE_{W_1*W_2}(\Phi_{\text{MMI}}) = \calE_{W_1*W_2}(\Phi_{\text{ML}}) \notag \\ 
&< \min\sth{\calE_{W_1}(\Phi_{\text{ML}}), \calE_{W_2}(\Phi_{\text{ML}})} \notag \\
&= \calE_W(\Phi_{\text{ML}}).
\end{align}
\end{IEEEproof}

Suboptimality of pairwise MMI is due to the rigidity of the scheme, specifically in not considering that images $\bfY$ and $\bfZ$ are dependent conditioned on $\bfX$. Thus, it behooves us to design a universal registration algorithm that does take such dependencies into account.

\subsection{Max Multiinformation for Multi-image Registration}

To consider all correlations across images, we use the multiinformation functional \cite{StudenyV1998} which is a more inclusive formulation of the underlying information in the system.
\begin{defn} \label{defn:multi_info}
The \emph{multiinformation} of random variables $X_1,\dots,X_n$ is defined as
\begin{equation} \label{eqn:multi_info}
I_M(X_1;\dots;X_n) = \sum_{i=1}^n H(X_i) - H(X_1,\dots,X_n).
\end{equation}
\end{defn} 
The chain rule for multiinformation is given by
\begin{equation}\label{eqn:multi_chain_rule}
I_M(X_1;\dots;X_n) = \sum_{i=2}^n I(X_i;X^{(i-1)}),
\end{equation}
where $X^{(i)} = (X_1,\dots,X_i)$. 

Let us define the multi-image registration method, which we refer to as the MM estimate, defined for three images as
\begin{equation}\label{eqn:MM_regn}
(\hat{\pi}_{Y,\text{MM}},\hat{\pi}_{Z,\text{MM}}) = \arg\max_{(\pi_1,\pi_2) \in \Pi^2} \hat{I}_M(X;Y_{\pi_1};Z_{\pi_2}),
\end{equation}
where $\hat{I}_M(\cdot)$ is the empirical estimate of multiinformation. The estimator can correspondingly be generalized to  any finite number of images. Note that the estimates obtained are the same as that of the pairwise MMI method, when $\bfY$ and $\bfZ$ are conditionally independent (empirically) given $\bfX$.

\begin{lemma} \label{lemma:MM_exp_consist}
The MM estimates are exponentially consistent for multi-image registration.
\end{lemma}
\begin{IEEEproof}
The proof is analogous to that of Theorem \ref{thm:MMI_ML_consistency}, and follows from the union bound and Theorem \ref{thm:ML_info_conv}.
\end{IEEEproof}

We now show that the MM method is asymptotically optimal in the error exponent using a type counting argument.

\subsection{Error Analysis}

We compare the error exponent of MM to that of ML. We again show the error exponent is characterized by the pair of transformations of $\bfY,\bfZ$ that are the hardest to differentiate.

Let $\bar{\pi},\bar{\pi}' \in \Pi^2$ and let $\psi(\bar{\pi},\bar{\pi}')$ be the binary hypothesis test of the problem of three-image registration when the set of permitted transformations is restricted to $\{\bar{\pi},\bar{\pi}'\}$. Let $P_{\bar{\pi},\bar{\pi}'}(\Phi), \calE_{\bar{\pi},\bar{\pi}'}(\Phi)$ be the error probability and the error exponent of $\Phi$ respectively.
\begin{lemma}\label{lemma:multi_to_bin_exp}
Let $\Phi$ be an asymptotically exponentially consistent estimator of three-image registration. Then,
\begin{equation}
\calE(\Phi) = \min_{\bar{\pi},\bar{\pi}' \in \Pi^2} \calE_{\bar{\pi},\bar{\pi}'}(\Phi).
\end{equation}
\end{lemma}
\begin{IEEEproof}
The lower bound on the error exponent follows directly from the union bound. Thus, $\calE(\Phi) \geq \min_{\bar{\pi},\bar{\pi}'} \calE_{\bar{\pi},\bar{\pi}'}(\Phi)$.

On the other hand, we have
\begin{align*}
P_e(\Phi) &= \frac{1}{|\Pi|^2} \sum_{\bar{\pi} \in \Pi} \sum_{\bar{\pi}'\in \Pi^2\backslash{\bar{\pi}}} \prob{\hat{\pi} = \bar{\pi}' \vert \bar{\pi}^* = \bar{\pi}} \\
&\geq \frac{1}{|\Pi|^2} \max_{\bar{\pi},\bar{\pi}' \in \Pi^2}P_{\bar{\pi},\bar{\pi}'}(\Phi).
\end{align*}
\end{IEEEproof}

Let $\bar{\pi}_0 = (\pi_0,\pi_0)$ correspond to the scenario where the image is not transformed.
\begin{lemma}\label{lemma:diff_equivalence_base}
For any $\bar{\pi}_1,\bar{\pi}_2 \in \Pi^2$, with $\bar{\pi}_i = (\pi_i,\pi_i')$, there exists $\bar{\pi} = (\pi,\pi)$ such that $\calE_{\bar{\pi}_1,\bar{\pi}_2}(\Phi) \geq \calE_{\bar{\pi}_0,\bar{\pi}}(\Phi)$, for $\Phi \in \{\Phi_{\text{ML}}, \Phi_{\text{MM}}\}$.
\end{lemma}
\begin{IEEEproof}
We prove the result for the ML decoder; the proof directly extends to the MM decoder.

Let $\tilde{\pi}_1$ be the permutation such that $\tilde{\pi}_1 \circ \pi_1' = \pi_1$. That is, the application of the transformation $\tilde{\pi}_1$ to an image that has been transformed by $\pi_1'$ results in an image that is effectively transformed by $\pi_1$. Let $\tilde{\pi}_2 = \tilde{\pi}_1 \circ \pi_2'$.

To obtain the ML decision for $\psi(\bar{\pi}_1,\bar{\pi}_2)$, let
\begin{align*}
(\hat{\pi}_Y,\hat{\pi}_Z) &= \arg\max_{(\pi_Y,\pi_Z) \in \{(\pi_1,\pi_1),(\pi_2,\tilde{\pi}_2)\}} W(\bfY_{\pi_Y},\bfZ'_{\pi_Z} \vert \bfX),
\end{align*}
where $\bfZ' = \bfZ_{\tilde{\pi}_1}$, is the received image, transformed by $\tilde{\pi}_1$. Then, $(\hat{\pi}_{Y,\text{ML}},\hat{\pi}_{Z,\text{ML}}) = (\hat{\pi}_Y,\tilde{\pi}_1^{-1} \circ \hat{\pi}_Z)$.

Since the source is i.i.d.\ and the channel is memoryless, $\calE_{\bar{\pi}_1,\bar{\pi}_2}(\Phi_{\text{ML}}) = \calE_{(\pi_1,\pi_1),(\pi_2,\tilde{\pi}_2)}(\Phi_{\text{ML}})$. Finally, if $\pi_1^{-1}$ is the inverse $\pi_1$, let $\pi = \pi_1^{-1} \circ \pi_2,\pi' = \pi_1^{-1} \circ \tilde{\pi}_2$. Then, 
\[
(\hat{\pi}_Y,\hat{\pi}_Z) = \arg\max_{(\pi_Y,\pi_Z) \in \{(\pi_0,\pi_0),(\pi,\pi')\}} W(\bfY'_{\pi_Y},\bfZ'_{\pi_Z} \vert \bfX),
\]
where $\bfY' = \bfY_{\pi_1^{-1}}$ and $\bfZ' = \bfZ_{\pi_1^{-1} \circ \tilde{\pi}_1}$, then $(\hat{\pi}_{Y,\text{ML}},\hat{\pi}_{Z,\text{ML}}) = (\pi_1 \circ \hat{\pi}_Y, (\pi_1^{-1} \circ \tilde{\pi}_1)^{-1} \circ \hat{\pi}_Z)$.

We note now that
\begin{equation}
D(P_{(\pi_0,\pi_0)}(X,Y,Z) \| P_{(\pi,\pi')}(X,Y,Z)) = I_M(X;Y;Z).
\end{equation}
Alternately,
\begin{equation}
D(P_{(\pi_0,\pi_0)}(X,Y,Z) \| P_{(\pi,\pi)}(X,Y,Z)) = I(X;Y,Z).
\end{equation}
From (\ref{eqn:multi_chain_rule}), it is evident that the binary hypothesis test in the second scenario is harder than the first. That is, the task of identifying if sequences are scrambled is easier if they are scrambled by different transformations than by the same. 

As the sequence of transformations are deterministic, the source i.i.d.\ and the channel memoryless, the result follows. The same arguments hold for the MM decoder.
\end{IEEEproof}

Lem.~\ref{lemma:diff_equivalence_base} implies that to study the error exponent of the multi-image registration problem, it suffices to study the error exponents of the binary hypothesis tests of the form $\psi(\bar{\pi}_0,\bar{\pi})$, for all $\pi\in\Pi$. Now we analyze error exponents of $\psi(\bar{\pi}_0,\bar{\pi})$.
\begin{theorem}\label{thm:multi_img_prob_bin}
Let $\bar{\pi} \in \Pi^2$. Then,
\begin{equation}\label{eqn:multi_img_prob_bin}
\lim_{n\rightarrow\infty} \frac{P_{\bar{\pi}_0,\bar{\pi}}(\Phi_{\text{MM}})}{P_{\bar{\pi}_0,\bar{\pi}}(\Phi_{\text{ML}})} = 1.
\end{equation}
\end{theorem}
\begin{IEEEproof}
Let $\bar{\pi} = (\pi,\pi)$. Using analysis similar to that of Thm.~\ref{thm:error_conv_binary}, we have
\begin{align*}
P_{\bar{\pi}_0,\bar{\pi}}(\Phi_{\text{MM}}) &= \sum_{q}\prod_{a\in\calX}\prod_{b\in\calY}\prod_{c\in\calZ}\pth{\prob{a,b,c}}^{nq(a,b,c)}\nu_{\text{MM}}(q) \\
&= P_{\bar{\pi}_1,\bar{\pi}_2}(\Phi_{\text{ML}}) \max_{q}\sth{\frac{\nu_{\text{MM}}(q)}{\nu_{\text{ML}}(q)}},
\end{align*}
where the sum is over the set of all possible joint types with 
\begin{align*}
\nu_{\text{MM}}(q) &= \sum_{\bfy,\bfz} \sum |T_{X \vert Y_{\pi_0},Z_{\pi_0},Y_{\pi},Z_{\pi}}^n(\bfy,\bfz)| \\
&= \sum |T_{Y_{\pi_0},Z_{\pi_0},Y_{\pi},Z_{\pi}}^n| \sum |T_{X \vert Y_{\pi_0},Z_{\pi_0},Y_{\pi},Z_{\pi}}^n|,
\end{align*}
where the first sum is taken over the set of all types $T_{Y_{\pi_0},Z_{\pi_0},Y_{\pi},Z_{\pi}}^n \subseteq T_{Y,Z}^n$ and the second sum is over the set of all conditional types $T_{X \vert Y_{\pi_0},Z_{\pi_0},Y_{\pi},Z_{\pi}}^n \subseteq T_{X \vert Y,Z}^n$ such that $\Phi_{\text{MM}}$ gives a decision error. Similarly, $\nu_{\text{ML}}(q)$ is defined analogously for ML decision errors.

The result follows from the forthcoming Lem.~\ref{lemma:number_ratio}.
\end{IEEEproof}

\begin{lemma}\label{lemma:number_ratio}
There exists a non-negative sequence $\sth{\epsilon_n}_{n \geq 1}$ with $\lim_{n \to \infty} n \epsilon_n =~0$, such that
\[
\max_{q}\sth{\frac{\nu_{\text{MM}}(q)}{\nu_{\text{ML}}(q)}} \leq 2^{n\epsilon_n}.
\]
\end{lemma}
\begin{IEEEproof}
MM is the same as minimizing joint entropy.
\begin{align*}
&\max_{(\pi_1,\pi_2) \in \Pi^2} \hat{I}(X;Y_{\pi_1};Z_{\pi_2}) \\
&\quad = \max_{(\pi_1,\pi_2)} \hat{H}(X) + \hat{H}(Y_{\pi_1}) + \hat{H}(Z_{\pi_2}) - \hat{H}(X,Y_{\pi_1},Z_{\pi_2}) \\
&\quad = \hat{H}(X) + \hat{H}(\tilde{Y}) + \hat{H}(\tilde{Z}) - \min_{(\pi_1,\pi_2) \in \Pi^2} \hat{H}(X,Y_{\pi_1},Z_{\pi_2}).
\end{align*}

Thus, for $V = (Y,Z)$, solving $\psi(\bar{\pi}_0,\bar{\pi})$ using MM is the same as minimizing joint empirical entropy $\hat{H}(X,V_{\hat{\pi}})$, over $\hat{\pi} \in \{\pi_0,\pi\}$. This is the same as the problem in Lem.~\ref{lemma:no_ratio} with a larger output alphabet. Thus, the result follows.
\end{IEEEproof}

\begin{theorem}\label{thm:three_img_exponent_match}
\begin{equation}
\calE(\Phi_{\text{MM}}) = \calE(\Phi_{\text{ML}}).
\end{equation}
\end{theorem}
\begin{IEEEproof}
The result follows from Lem.~\ref{lemma:multi_to_bin_exp} and {\ref{lemma:diff_equivalence_base}}, and Thm.~\ref{thm:multi_img_prob_bin}.
\end{IEEEproof}

These results directly extend to registering any finite number of images. This indicates that the MM method for multi-image registration is asymptotically optimal and universal.

\begin{remark}
The results for image models with i.i.d.\ pixels extend directly to memoryless or exchangeable sources. That is, we can generalize the proofs of asymptotic optimality of the MM method for image registration to image models where pixels are drawn from a memoryless or an exchangeable distribution. This follows as the probabilities of sequences with the same type are of equal probability for such sources.
\end{remark}

\section{Joint Clustering and Registration} \label{sec:joint_cluster_regn}

Now we return to the original problem of joint clustering and registration of multiple images. As mentioned in the image model, we know that images corresponding to different scenes are independent. We are interested in estimating the partition and registering images within each cluster.

The ML estimates are obtained as
\begin{equation} \label{eqn:ml_clust_regn}
\pth{\hat{P}_{\text{ML}}, \hat{\pi}_{\text{ML}}} = \arg\max_{P \in \calP, \bar{\pi} \in \Pi^{m}} \prod_{C \in P} \prod_{i=1}^n \prob{\bfX_{\pi^{(C)}(i)}^{(C)}},
\end{equation}
where the probability is computed by averaging over scene configurations and the corresponding channel model. 

However, in this case it is worth noting that the ML estimates are not Bayes optimal as the prior on the set of possible permutations is not uniform. For instance, if $p_i = \frac{1}{\ell}$ for all $i \in [\ell]$, then for any partition $P \in \calP$,
\[
\prob{P^* = P} = \frac{m!}{k!} \frac{1}{\ell^m},
\]
where $k$ is the number of clusters in the partition $P$. Hence, in this context, the Bayes optimal test is the likelihood ratio test, i.e., the maximum a posteriori (MAP) estimate $\Phi_{\text{MAP}}$ given by
\begin{align}
\pth{\hat{P}_{\text{MAP}}, \hat{\pi}_{\text{MAP}}} &= \arg\max_{P, \bar{\pi}} \prob{P^* = P} \prod_{C \in P} \prod_{i=1}^n \prob{\bfX_{\pi^{(C)}(i)}^{(C)}},  \label{eqn:map_clust_regn}
\end{align}

Additionally, both ML and MAP estimates require knowledge of channel and prior distributions, and are also hard to compute. Hence we focus on the design of computationally efficient and exponentially consistent algorithms.

\subsection{Clustering Criteria}

Designing any unsupervised clustering algorithm requires a criterion to quantify similarity. So algorithms are designed in contexts such as knowing the number of clusters or under a notion of closeness for objects of the same cluster.

Here, we know that the similarity criterion is dependence of pixel values among images of the same cluster. To address this, one could adopt any of the following methods.
\begin{enumerate}
\item[(B1)] {\it $\epsilon$-likeness:} A given source and channel model for images, $\bfX^{[m]}$, is said to satisfy \emph{$\epsilon$-likeness} criterion if 
\[
\min_{P^* \in \calP} \min_{P \not\preceq P^*, P \not\succeq P^*} I_C^{P^*}(X^{[m]}) - I_C^P(X^{[m]}) \geq \epsilon.
\] 
\item[(B2)] {\it Given number of clusters:} Given the number of clusters $k$ in the set of images, we can define an exponentially consistent universal clustering algorithm.
\item[(B3)] {\it Non-exponentially consistent:} Any two images $\bfX,\bfY$ that belong to different clusters are independent, i.e., $I(X;Y) = 0$. Hence, using a threshold $\gamma_n$, decreasing with $n$, we can define a consistent clustering algorithm which however lacks exponential consistency, cf.~\cite{RamanV2017a}.
\item[(B4)] {\it Hierarchical clustering:} If it suffices to create the tree of similarity among images through hierarchical clustering, then we can define a consistent algorithm that determines such topological relation among images.
\end{enumerate} 

Criterion (B1) imposes a restriction on the allowed priors and channels $W$ and may be interpreted as a capacity. On the other hand (B2) restricts the space of partitions much like $k$-means clustering. Criterion (B3) focuses on the design of asymptotically consistent universal clustering algorithms albeit with sub-exponential consistency. Finally, criterion (B4) aims to develop a topology of independence-based similarity relations among images. We address the clustering problem for each of these criteria.

\subsection{Multivariate Information Functionals for Clustering}

Clustering random variables using information functionals has been well studied \cite{NaganoKI2010,ChanAEKL2015}. Since image clustering here is independence-based, we adopt the minimum partition information (MPI) framework.

\begin{defn} \label{defn:part_info}
For a set of random variables $\{Z_1,\dots,Z_n\}$ with $n>1$, and a partition $P$ of $[n]$ with $|P| > 1$, the \emph{partition information} is defined as
\begin{equation} \label{eqn:part_info}
I_P \pth{Z^{[n]}} =\frac{1}{|P|-1}\qth{ \pth{ \sum_{C \in P} H \pth{Z^{C}} }  - H\pth{Z^{[n]}} }.
\end{equation}
\end{defn}
The chain rule for partition information is given as 
\begin{equation} \label{eqn:chain_rule}
I_{P}(Z^{[n]}) =\frac{1}{k-1}\sum_{i=1}^{k-1}I\pth{Z^{C_i};Z^{\sth{\cup_{j=i+1}^k C_j}}} ,
\end{equation}
where $P=\{C_1,\dots,C_k\}$. For $P = \{ \{1\}, \dots, \{n\} \}$, i.e., each object constitutes its own cluster,
\[
I_{P}(Z^{[n]}) =\frac{1}{n-1}I_M(Z^{[n]}).
\]
The minimum residual independence in a collection of random variables is quantified by the MPI.
\begin{defn} \label{defn:MPI}
For a set of random variables $Z^{[n]}$ with $n>1$, the \emph{minimum partition information} is 
\begin{equation}
I_{\text{MPI}}(Z^{[n]})  = \min_{P \in \calP'} I_P (Z^{[n]}) ,
\end{equation}
where $\mathcal{P}'$ is the set of all partitions of $[n]$ such that for all $P \in \calP'$, $|P| > 1$.
\end{defn}

We note here that the partition that minimizes the partition information is not necessarily unique.
\begin{defn} \label{defn:fund_part}
Let $\tilde{\calP} = \arg\min_{P \in \calP'} I_P \pth{Z^{[n]}}$. Then, the \emph{fundamental partition} of the set of random variables $Z^{[n]}$ is the finest partition $P^* \in \tilde{\calP}$. That is, $P^* \in \tilde{\calP}$ such that for all $P \in \tilde{\calP}$, $P^* \preceq P$.
\end{defn}

For the problem at hand, let $X^{(i)}$ be the random variable representing a pixel of image $i$. Then, the images satisfy the following properties.
\begin{lemma}\label{lemma:fund_part_clust}
If the images are self-aligned, i.e., $\bar{\pi}^* = \bar{\pi}_0$, and there are at least two clusters, then the correct clustering is given by the fundamental partition and the MPI is zero. 
\end{lemma}
\begin{IEEEproof}
Let $P^*$ be the correct clustering. First we note that $I_{P^*}(X^{[m]}) = 0$, as images corresponding to different clusters are independent. Further we know that the partition information is non-negative. Hence, 
\[
P^* \in \arg\min_{P \in \calP'} I_P(X^{[m]}).
\]

Let $\tilde{\calP} = \{P' \in \calP':P' \succeq P^*\}$ and let $P \in \calP'\backslash{\tilde{\calP}}$. Then there exist clusters $C_1,C_2 \in P$ such that there exist images in each cluster corresponding to the same image. This indicates that $I(X^{C_1};X^{C_2}) > 0$. Thus, from (\ref{eqn:chain_rule}),
\[
I_{P^*}(X^{[m]}) \geq I(X^{C_1};X^{C_2}) > 0.
\]

This indicates that the set of all partitions that minimize the partition information is the set $\tilde{P}$. Hence the fundamental partition is the correct clustering.
\end{IEEEproof}

\begin{lemma}\label{lemma:fund_part_regn}
Let $\bar{\pi} = (\pi_1,\dots,\pi_m)$ be the estimated transformations and let $\bar{\pi}^*$ be the correct registration. Then, if $\bar{\pi} \neq \bar{\pi}^*$, then the fundamental partition $\hat{P}$ of $\{X^{(1)}_{\pi_1},\dots,X^{(m)}_{\pi_m}\}$ satisfies $\hat{P} \prec P^*$, where $P^*$ is the correct clustering.
\end{lemma}
\begin{IEEEproof}
We first note that images that correspond to different scenes are independent of each other, irrespective of the transformation. Second, an image that is incorrectly registered appears independent of any other image corresponding to the same scene. This in turn yields the result.
\end{IEEEproof}

These properties provide a natural estimator for joint clustering and registration, provided the information values can be computed accurately.
\begin{corollary} \label{cor:clust_regn_est}
Let $\hat{P}_{\bar{\pi}}$ be the fundamental partition corresponding to the estimated transformation vector $\bar{\pi}$. Then $(P^*,\bar{\pi}^*)$ is the densest partition in $\{\hat{P}_{\bar{\pi}}: \bar{\pi} \in \Pi^m\}$ and the corresponding transformation vector.
\end{corollary}
\begin{IEEEproof}
This follows directly from Lem.~\ref{lemma:fund_part_clust} and \ref{lemma:fund_part_regn}.
\end{IEEEproof}

It is worth noting that the partition information is submodular and constitutes a Dilworth truncation lattice through the residual entropy function \cite{ChanAEKL2015}. Thus, the fundamental partition may be obtained efficiently using submodular function minimization in polynomial time \cite{ChanAZKL2016}. We do not discuss algorithm implementation details here, since we focus on the consistency and the error exponents of algorithms.

We now simplify the clustering criterion for better understanding and ease of analysis. Let us define a multivariate information measure called \emph{cluster information} to quantify intra-cluster information.
\begin{defn} \label{defn:clust_info}
The \emph{cluster information} of $Z^{[n]}$ for partition $P = \{C_1,\dots,C_k\}$ of $[n]$ is
\begin{equation}\label{eqn:clust_info}
I_C^{(P)}(Z_1;\dots;Z_n) = \sum_{C \in P} I_M(Z^C).
\end{equation}
\end{defn} 
Thus for any $P \in \calP'$, and random variables $X^{[m]}$,
\begin{align}
I_P(X^{[m]}) = \frac{\pth{I_M(X_1;\dots;X_m) - I_C^{(P)}(X_1;\dots;X_m)}}{(|P|-1)}. \label{eqn:clust_part_info}
\end{align}
Thus, the fundamental partition maximizes the cluster information if the MPI is $0$.

As we seek to identify the partition that maximizes the cluster information, it is worthwhile understanding its computational complexity.
\begin{lemma}\label{lemma:supermodularity}
The clustering information of a set of random variables $\{Z_1,\dots,Z_n\}$ is supermodular.
\end{lemma}
\begin{IEEEproof}
The clustering information may be decomposed as
\[
I_C^{(P)}(Z_1;\dots;Z_n) = \sum_{i=1}^n H(Z_i) - \sum_{C \in P} H(Z^C).
\]
Since the entropy function is submodular and since the clustering information is the difference between a modular function and a submodular function, it is supermodular.
\end{IEEEproof}

Supermodular function maximization can be done efficiently in polynomial time. From Lem.~\ref{lemma:supermodularity} we see that given the distribution, we can obtain the fundamental partition efficiently.

We now use these observations to define clustering algorithms using plug-in estimates, under each clustering criterion.

\subsection{Consistency of Plug-in Estimates}

Since the multivariate information estimates are obtained empirically, we first show that the plug-in estimates are exponentially consistent.
\begin{lemma}\label{lemma:MPI_consist}
The plug-in estimates of partition and cluster information are exponentially consistent.
\end{lemma}
\begin{IEEEproof}
From the chain rule (\ref{eqn:chain_rule}) and the union bound, we know that for all $\epsilon>0$, there exists $\delta_\epsilon>0$ and constant $c$ such that
\begin{equation}\label{eqn:PI_conv}
\prob{\left|\hat{I}_P\pth{X^{[m]}} - I_P\pth{X^{[m]}}\right| > \epsilon} \leq c\exp\pth{-n\delta_\epsilon},
\end{equation}
from Theorem \ref{thm:ML_info_conv}. 

The result for the cluster information estimate follows from \eqref{eqn:clust_info}, the union bound, and the exponential consistency of plug-in estimates of multiinformation as
\begin{equation} \notag
\prob{\left| \hat{I}_C^{(P)}\pth{X^{[m]}} - I_C^{(P)}\pth{X^{[m]}} \right| > \epsilon} \leq C \exp\pth{ -n \delta_{\epsilon}},
\end{equation}
where $C = 2m$, $\delta_{\epsilon} = \tfrac{\epsilon^4}{32m^4 |\calX|^{2m} \log 2} + o(1) = \theta \epsilon^4 + o(1)$.
\end{IEEEproof}

We now use the plug-in estimates to define universal clustering algorithms.

\subsection{$\epsilon$-like clustering}

We define $\epsilon$-like clustering, Alg.~\ref{algo:clust_est}, using the fact that the cluster information is maximized by the correct clustering. 

\begin{algorithm}[t]
  \caption{$\epsilon$-like Clustering, $\Phi_{\text{C}}\pth{\bfX^{(1)},\dots,\bfX^{(m)},\epsilon}$}
  \label{algo:clust_est}
  \begin{algorithmic}[ht]
  	\FORALL {$\bar{\pi} \in \Pi^m$}
   		\STATE{Compute empirical pmf $\hat{p}\pth{X^{(1)}_{\pi_1},\dots,X^{(m)}_{\pi_m}}$}
   		\STATE{$\tilde{I} = \max_{P \in \calP} \hat{I}_C^{(P)}(X^{(1)}_{\pi_1},\dots,X^{(m)}_{\pi_m})$}
   		\STATE{$P_{\pi}~=~\text{Finest}\sth{P : \hat{I}_C^{(P)}(X^{(1)}_{\pi_1},\dots,X^{(m)}_{\pi_m}) \geq \tilde{I} - \tfrac{\epsilon}{2}}$, where $\text{Finest}\sth{\cdot}$ refers to the finest partition in the set.}
   	\ENDFOR
   	\STATE{$(\hat{P},\hat{\pi}) = \sth{(P,\bar{\pi}): P = \hat{P}_{\bar{\pi}} \succeq \hat{P}_{\bar{\pi}'}, \text{ for all } \bar{\pi}' \neq \bar{\pi}}$}
  \end{algorithmic}
 \end{algorithm}

\begin{lemma}\label{lemma:clust_info_consist}
Let the source and channel be $\epsilon$-like, for some $\epsilon>0$. Then, $\Phi_C$ is exponentially consistent for joint clustering and registration.
\end{lemma}
\begin{IEEEproof}
First let the estimated transformation be correct, $\hat{\pi} = \pi^*$. Let $\tilde{I} = \max_{P \in \calP} \hat{I}_C^{(P)}(X^{(1)}_{\pi_1},\dots,X^{(m)}_{\pi_m})$, $\tilde{P}$ the maximizing partition. Then, there are constants $c,\delta_{\epsilon}$ such that
\begin{align}
&\prob{\hat{I}_C^{(P^*)}(X^{(1)}_{\pi_1},\dots,X^{(m)}_{\pi_1}) \leq \tilde{I} - \tfrac{\epsilon}{2}} \notag \\ 
&\quad \leq \prob{|\tilde{I} - \hat{I}_C^{(P^*)}(X^{(1)}_{\pi_1},\dots,X^{(m)}_{\pi_m})| \geq \tfrac{\epsilon}{2}} \notag\\
&\quad \leq \prob{|\hat{I}_C^{(\tilde{P})} - I_C^{(\tilde{P})}| + |\hat{I}_C^{(P^*)} - I_C^{(P^*)}| \geq \tfrac{\epsilon}{2}} \label{eqn:triang_ineq}\\
&\quad \leq 2c\exp\pth{-n\delta_{\epsilon/4}}, \label{eqn:union_exp_const}
\end{align} 
where \eqref{eqn:triang_ineq} follows from the triangle inequality and the fact that the multiinformation is maximized by the correct clustering, and \eqref{eqn:union_exp_const} follows from the union bound and Lem.~\ref{lemma:MPI_consist}. Here in \eqref{eqn:triang_ineq} the information measures are computed for the random variables $X^{[[m]}_{\bar{\pi}} \triangleq (X^{(1)}_{\pi_1},\dots,X^{(m)}_{\pi_m})$.

Further, for any $P \in \calP$ such that $P \not\succeq P^*$ and $P\not\preceq P^*$
\begin{align}
&\prob{\hat{I}_C^{(P)}(X^{(1)}_{\pi_1},\dots,X^{(m)}_{\pi_1}) \geq \tilde{I} - \tfrac{\epsilon}{2}} \notag \\
&\quad \leq \prob{\hat{I}_C^{(P^*)}(X^{(1)}_{\pi_1},\dots,X^{(m)}_{\pi_m}) - \hat{I}_C^{(P)}(X^{(1)}_{\pi_1},\dots,X^{(m)}_{\pi_m}) \leq \tfrac{\epsilon}{2}} \label{eqn:emp_max} \\
&\quad \leq \prob{\left|\pth{\hat{I}_C^{(P^*)} - I_C^{(P^*)}} - \pth{\hat{I}_C^{(P)} - I_C^{(P)}} \right| \geq \tfrac{\epsilon}{2}} \label{eqn:info_diff}\\
&\quad \leq 2c\exp\pth{-n\delta_{\epsilon/4}}, \label{eqn:tri_union_exp_const}
\end{align}
where \eqref{eqn:emp_max} follows from the fact that $\tilde{I}$ is the maximum empirical cluster information and \eqref{eqn:info_diff} follows from the $\epsilon$-likeness criterion. Finally, \eqref{eqn:tri_union_exp_const} follows from the triangle inequality, union bound and Lem.~\ref{lemma:MPI_consist}. Here again, in \eqref{eqn:info_diff} the information measures are computed for $X^{[[m]}_{\bar{\pi}}$.

From \eqref{eqn:union_exp_const}, \eqref{eqn:tri_union_exp_const}, and the union bound we know that
\begin{equation} \label{eqn:non_exist}
\prob{\hat{P}_{\bar{\pi}} \neq P^*} \leq 4c\exp\pth{-n\delta_{\epsilon/4}}.
\end{equation}

Now, invoking Lem.~\ref{lemma:fund_part_regn}, we know from similar analysis, that the densest fundamental partition is exactly $P^*$. More specifically, for any $\bar{\pi} \neq \bar{\pi}^*$, the equivalent fundamental partition is finer than $P^*$. This in turn indicates that
\begin{equation}
P_e(\Phi_C) \leq 4c|\Pi| \exp\pth{-n\delta_{\epsilon/4}},
\end{equation}
owing to the union bound and \eqref{eqn:non_exist}.
\end{IEEEproof}

From \eqref{eqn:clust_part_info}, we can additionally see that an equivalent clustering algorithm can be defined in terms of the MPI functional and that it is also exponentially consistent, provided the underlying partition has at least two clusters.

\subsection{$K$-info clustering}

Under (B2), i.e., given number of clusters $K$ in the set of images, let $\calP_K \subset \calP$ be the set of all partitions consisting $K$ clusters. Then, much in the spirit of $K$-means clustering, we define the $K$-info clustering estimate as
\begin{equation}
(\hat{P},\hat{\pi}) = \arg\max_{P \in \calP_K, \bar{\pi} \in {\Pi}^{m}} \hat{I}_C^{(P)}(X^{(1)}_{\pi_1},\dots,X^{(m)}_{\pi_1}) 
\end{equation}
Again, this can be extended directly to use the MPI functional.

\begin{lemma}\label{lemma:clust_K_info_consist}
Given the number of clusters $K$ in the set, the $K$-info clustering estimates are exponentially consistent.
\end{lemma}
\begin{IEEEproof}
Let $P_{\bar{\pi}} = \arg\max_{P \in \calP_K} I_C^{(P)}(X^{[m]}_{\bar{\pi}})$ and $\hat{P}_{\bar{\pi}} = \arg\max_{P \in \calP_K} \hat{I}_C^{(P)}(X^{[m]}_{\bar{\pi}})$. Then, for any $\bar{\pi} \in \Pi^m$,
\begin{align}
\prob{\hat{P}_{\bar{\pi}} \neq P_{\bar{\pi}}} &= \prob{\hat{I}_C^{(\hat{P}_{\bar{\pi}})}(X^{[m]}_{\bar{\pi}}) > I_C^{(P_{\bar{\pi}})}(X^{[m]}_{\bar{\pi}})} \notag \\
&\leq 2(|\calP_K| - 1)c \exp\pth{-n\delta_{\epsilon_{\bar{\pi}}}}, \label{eqn:union_info_consist}
\end{align}
where \eqref{eqn:union_info_consist} follows from the union bound and Lem.~\ref{lemma:MPI_consist}, for $\epsilon_{\bar{\pi}} = I_C^{(P_{\bar{\pi}})}(X^{[m]}_{\bar{\pi}}) - \max_{P \neq P_{\bar{\pi}}} I_C^{(P)}(X^{[m]}_{\bar{\pi}})$.

Next, for any $\bar{\pi} \neq \bar{\pi}^*$, we have
\begin{align}
&\prob{\hat{I}_C^{(P_{\bar{\pi}})}(X^{[m]}_{\bar{\pi}}) > \hat{I}_C^{(P_{\bar{\pi}^*})}(X^{[m]}_{\bar{\pi}^*})} \notag \\
&\quad \leq \prob{\left|\hat{I}_C^{(P_{\bar{\pi}})} - I_C^{(P_{\bar{\pi}})} - \hat{I}_C^{(P_{\bar{\pi}^*})} + I_C^{(P_{\bar{\pi}^*})} \right| \geq I_C^{(P_{\bar{\pi}^*})} - I_C^{(P_{\bar{\pi}})}} \label{eqn:info_shorts}\\
&\quad \leq 2c\exp\pth{-n\delta_{\tilde{\epsilon}_{\bar{\pi}}}} \label{eqn:pi_side_exp},
\end{align}
where $\tilde{\epsilon}_{\bar{\pi}} = I_C^{(P_{\bar{\pi}^*})}(X^{[m]}_{\bar{\pi}^*}) - I_C^{(P_{\bar{\pi}})}(X^{[m]}_{\bar{\pi}})$. In \eqref{eqn:info_shorts}, $I_C^{(P_{\bar{\pi}^*})}, \hat{I}_C^{(P_{\bar{\pi}^*})}$ are the information measures computed for $X^{[m]}_{\bar{\pi}^*}$, while $I_C^{(P_{\bar{\pi}})}, \hat{I}_C^{(P_{\bar{\pi}})}$ are computed for $X^{[m]}_{\bar{\pi}}$.

Finally, the result follows from \eqref{eqn:union_info_consist}, \eqref{eqn:pi_side_exp}, the union bound, and the fact that $|\Pi| = O(n^{\alpha})$.
\end{IEEEproof}

\subsection{Clustering with sub-exponential consistency}

Under criterion (B3) we seek a universal clustering algorithm that is not aware of the underlying parameters of $\epsilon$-likeness and number of clusters $K$. 

We know that independence characterizes dissimilarity of images. Given a source and channel model, there exists $\epsilon > 0$ such that the set of images are $\epsilon$-alike. However, the value of $\epsilon$ is unknown to the decoder. Since the plug-in estimates are exponentially consistent, we adapt Alg.~\ref{algo:clust_est} to work with a dynamic threshold that takes into consideration the resolution (the number of pixels) of the image. The threshold-based clustering algorithm is given in Alg.~\ref{algo:thresh_clust}.

\begin{algorithm}[t]
  \caption{Thresholded Clustering, $\Phi_{\text{T}}\pth{\bfX^{(1)},\dots,\bfX^{(m)},\alpha}$}
  \label{algo:thresh_clust}
  \begin{algorithmic}[t]
  \STATE{$\gamma_n \leftarrow c_1 n^{-\alpha}$, for some constant $c_1 > 0$}
  	\FORALL {$\bar{\pi} \in \Pi^m$}
   		\STATE{Compute empirical pmf $\hat{p}\pth{X^{(1)}_{\pi_1},\dots,X^{(m)}_{\pi_m}}$}
   		\STATE{$\tilde{I} = \max_{P \in \calP} \hat{I}_C^{(P)}(X^{(1)}_{\pi_1},\dots,X^{(m)}_{\pi_m})$}
   		\STATE{$P_{\pi} = \text{Finest}\sth{P : \hat{I}_C^{(P)}(X^{(1)}_{\pi_1},\dots,X^{(m)}_{\pi_m}) \geq \tilde{I} - \gamma_n}$}
   	\ENDFOR
   	\STATE{$(\hat{P},\hat{\pi}) = \sth{(P,\bar{\pi}): P = \hat{P}_{\bar{\pi}} \succeq \hat{P}_{\bar{\pi}'}, \text{ for all } \bar{\pi}' \neq \bar{\pi}}$}
  \end{algorithmic}
 \end{algorithm}

\begin{lemma}
Let $\alpha \in (0,\tfrac{1}{4})$. Then, the thresholded clustering algorithm, $\Phi_T$ is asymptotically consistent.
\end{lemma}
\begin{IEEEproof}
Let the given source and channel be $\epsilon$-like. Since $\gamma_n$ is decreasing with $n$, there exists $N_{\epsilon} < \infty$ such that for all $n > N_{\epsilon}$, $\gamma_n < \epsilon$. The proof now follows analogous to the proof of Lem.~\ref{lemma:clust_info_consist}, by observation $n \delta_{\gamma_n} \to \infty$ as $n \to \infty$.
\end{IEEEproof}

We see that at the expense of exponential consistency, there is a reliable universal clustering scheme that only uses independence for clustering and registering images.

\subsection{Hierarchical Clustering}

Finally we consider clustering according to criterion (B4). Here, we aim to establish the hierarchical clustering relation among images to establish a dendrogram for the images. From the nature of independence among dissimilar images, we design a natural algorithm for hierarchical clustering, Alg.~\ref{algo:hierarch_clust}.

\begin{algorithm}[t]
  \caption{Hierarchical Clustering, $\Phi_{\text{H}}\pth{\bfX^{(1)},\dots,\bfX^{(m)}}$}
  \label{algo:hierarch_clust}
  \begin{algorithmic}[t]
  \STATE{$\hat{\pi} = \arg\max_{\bar{\pi} \in \Pi^m} \hat{I}_M(X^{[m]}_{\bar{\pi}})$}
  \STATE{$\bfY^{[m]} \leftarrow X^{[m]}_{\hat{\pi}}$}
  \STATE{$P \leftarrow \{[m]\}$, $\hat{P}(1) \leftarrow P$}
  \FOR{$k=2$ \TO $m$}
  	\FORALL{$C \in P$}
  		\STATE{$J_C \leftarrow \max_{\tilde{P} \in \calP_2(C)} I_C^{(\tilde{P})}(Y^C)$}
  		\STATE{$\tilde{P}_C \leftarrow \arg\max_{\tilde{P} \in \calP_2(C)} I_C^{(\tilde{P})}(Y^C)$}
  	\ENDFOR
  	\STATE{$\tilde{C} \leftarrow \arg\max_{C \in P} J_C$}
  	\STATE{$P \leftarrow P \backslash \tilde{C}$}
  	\STATE{$P \leftarrow P \cup \tilde{P}_{\tilde{C}}$}
  	\STATE{$\hat{P}(k) \leftarrow P$}
  \ENDFOR
  \end{algorithmic}
 \end{algorithm}

Fundamentally, at each stage, the algorithm splits into two, one cluster from the existing partition. More specifically, the cluster that has the most impact upon such a split in terms of the cluster information is chosen at each stage.

\begin{lemma}\label{lemma:hierarch_consist}
Consider a set of images with $K$ clusters. Then, the resulting partition at iteration $k=K$ of $\Phi_H$ is almost surely the correct clustering. 

Alternately, if the system is $\epsilon$-like, then for sufficiently large $n$, at iteration $k = |P^*|$ the estimated partition is correct with arbitrarily high probability. Further, the transformation estimates are exponentially consistent.
\end{lemma}
\begin{IEEEproof}
The proof follows directly from exponential consistency of the other clustering and registration algorithms.
\end{IEEEproof}

\subsection{Computational and Sample Complexity}

Although not the principal focus, we briefly note the computational and sample complexity of the algorithms. 

We have established that the fundamental partitions for cluster information may be obtained efficiently for any chosen transformation vector, given the joint distribution. However, exploring the neighborhood of the maximizing partition is a harder problem to address and it is not clear if there is an efficient algorithm, other than exhaustive search, to do this.

Further, identifying the correct transformation involves an exhaustive search in the absence of additional information on transformation nature, dependency in the transformations across different images, or underlying correlations across pixels of an image. More specifically, since $|\Pi| = O(n^{\alpha})$, the complexity of exhaustive search is $O(n^{\alpha m})$, i.e., exponential in the number of images. For a constant number of images to be clustered and registered, this method is efficient.

With regard to sample complexity, as we deal with the computation of information measures across a set of $m$ random variables, to suppress the bias in the plug-in estimates, the sample complexity of the clustering and registration algorithms is $O(r^m)$. This translates to the requirement of high resolution images. Additional information on structure in the underlying image model could be used to reduce the sample complexity.

\section{Large-scale Registration of Images}

Now consider the problem of registering a large set of $m$ images with a limited number of $n$ pixels (resolution) each. We analyze the scenario where the number of pixels in each image is sub-exponential in the number of images, i.e., $n = o\pth{\tfrac{r^m}{m}}$. Then, accurate computation of the multiinformation of all images is not universally feasible from the lower bounds in entropy estimation established in \cite{ValiantV2011}. 

Consider a set of corrupted and transformed images $\bfY^{(1)},\dots,\bfY^{(m)}$. Even though MM is asymptotically optimal for image registration, it is not feasible due to limited availability of pixels. We thus provide necessary and sufficient conditions on the number of pixels for a given number of images for consistent, universal, multi-image registration.

\subsection{Necessary Conditions}

We first establish necessary conditions on the number of pixels for consistent image registration. Note that registering images is equivalent to clustering them by the transformations. We use ideas from \cite{RamanV2017a} to lower bound registration cost.

\begin{theorem} \label{thm:necessary_cost}
Given a set of $m$ copies of an image, there exists a channel-aware registration algorithm only if $n = \Omega(\log m)$.
\end{theorem}
\begin{IEEEproof}
Consider the simpler case where only two transformations exist, $\Pi = \{\pi_0, \pi\}$. Group the set of images into pairs at random, and let $\psi = \{(i,j): i,j \in [m]\}$ be the set of pairs. For ease, assume that $m$ is divisible by $2$. Define the binary hypothesis tests given by
\[
\psi_{ij}: \begin{cases}
H_0 : \pi_i = \pi_j \\
H_1 : \pi_i \neq \pi_j
\end{cases}
\]
Then, from the Kailath lower bound \cite{Kailath1967}, we have
\begin{equation} \label{eqn:Kailath_lb}
\prob{\text{Error in } \psi_{ij}} \geq \exp\pth{-\tfrac{1}{2} (D(p_0 \| p_1) + D(p_1 \| p_0))},
\end{equation}
where $p_i$ is the conditional distribution of the observation, given hypothesis $H_i$. Let $\Delta_{i,j} = \tfrac{1}{2} (D(p_0 \| p_1) + D(p_1 \| p_0))$ for each $(i,j) \in \psi$, and let $\Delta_{\text{max}} = \max_{(i,j) \in \psi} \Delta_{i,j}$.

Here, observations are transformed images; under $H_0$,
\[
p_0(\bfY^{(i)},\bfY^{(j)}) = \prod_{k=1}^n \prob{Y^{(i)}_k} \tilde{W}(Y^{(j)}_k \vert Y^{(i)}_k),
\]
where $\tilde{W}$ is the equivalent channel relating image $\bfY^{(i)}$ to $\bfY^{(j)}$; $\tilde{W}$ is specific to the pair $(i,j)$ and we have ignored the indices for simplicity. Similarly, under hypothesis $H_1$ 
\begin{align*}
&p_1(\bfY^{(i)},\bfY^{(j)}) = \frac{1}{2}\prod_{k=1}^n \prob{Y^{(i)}_{\pi(k)}} \tilde{W}\pth{Y^{(j)}_k \vert Y^{(i)}_{\pi(k)}} \\
&\quad + \frac{1}{2} \prod_{k=1}^n \prob{Y^{(i)}_{\pi^{-1}(k)}} \tilde{W}\pth{Y^{(j)}_k \vert Y^{(i)}_{\pi^{-1}(k)}}.
\end{align*}
Let $q(\bfY^{(i)},\bfY^{(j)}) = \prod_{k=1}^n \prob{Y^{(i)}_{\pi(k)}} \tilde{W}\pth{Y^{(j)}_k \vert Y^{(i)}_{\pi(k)}}$. Then, 
\begin{align}
&D(p_0 \| q) = \sum_{k=1}^n D\pth{\tilde{W}(Y \vert X_{\pi(k)}) \| \tilde{W}(Y \vert X_k) \vert X_k, X_{\pi(k)}} \notag \\
&\quad = (n - |\calI_{\pi}|) D(\tilde{W}(Y \vert X_1) \| \tilde{W}(Y \vert X_2) \vert X_1, X_2), \label{eqn:D_pq_comp}
\end{align}
where $X_1,X_2$ are i.i.d.\ copies of pixels corresponding to image $i$. Here \eqref{eqn:D_pq_comp} holds as only displaced pixels contribute non-zero distance values and the i.i.d.\ nature of drawing pixels. From \eqref{eqn:channel_dist_bd}, using the fact that $|\calI_{\pi}| = o(n)$ in \eqref{eqn:D_pq_comp}, the symmetry in the distributions, and convexity of KL divergence, we get
\begin{equation} \label{eqn:delta_bound}
\Delta_{i,j} \leq (n - o(n)) \theta_{M}, \quad \text{for all } (i,j) \in \psi
\end{equation}

Correct registration implies correct inference in all the hypothesis tests $\psi_{ij}$. Thus, for any registration algorithm $\Phi$:
\begin{align}
P_e(\Phi) &\geq 1 - \prod_{(i,j) \in \psi}\pth{1 - \prob{\text{Error in } \psi_{ij}}} \notag \\
&\geq 1 - \pth{1 - \exp\pth{-\Delta_{\text{max}}}}^{m/2} \label{eqn:use_Kailath} \\
&\geq 1 - \pth{1- \exp\pth{-(n-o(n)) \theta_{M}}}^{m/2}, \label{eqn:regn_error_lb}
\end{align}
where \eqref{eqn:use_Kailath} follows from \eqref{eqn:Kailath_lb}, and \eqref{eqn:regn_error_lb} follows from \eqref{eqn:delta_bound}.

Finally, from \eqref{eqn:regn_error_lb}, we note that the registration algorithm is consistent with $m \to \infty$ only if $n = \Omega(\log m)$.
\end{IEEEproof}

\subsection{Achievable Scheme}

Having obtained necessary conditions for large-scale image registration, we now provide an information-based registration algorithm that is order-optimal in the number of pixels.

We know MM is asymptotically optimal for a fixed and finite number of images. Thus, we can register the images by splitting them into subsets of appropriate size, keeping one reference common in all sets, Alg.~\ref{algo:block_reg}.

\begin{algorithm}[t]
  \caption{Blockwise Registration}
  \label{algo:block_reg}
  \begin{algorithmic}[t]
  \STATE{$k \leftarrow \arg\min_{2 \leq \ell  \leq (\log n)/(\log r)} \ell \log |\Pi| - \frac{nc}{r^{\ell}} {\ell}^4 (\log r)^4 - \log (\ell-1)$, where $c$ is a sufficiently small constant}
  \FOR {$i = 1$ to $\lceil \frac{m-1}{k - 1} \rceil$}
  	\STATE{$T_i \leftarrow \sth{(i-1)*(k-1)+1,\dots,i*(k-1)} \cup \{m\}$}
  	\STATE{Determine $\{\hat{\pi}_j : j\in T_i\}$ using MM method}
  \ENDFOR	
  \end{algorithmic}
 \end{algorithm}

The size of the subsets of images is essentially chosen to be a constant and for $k = 2$ reduces to pairwise MMI.

\begin{theorem} \label{thm:sufficient_cost}
For any channel $W$, for sufficiently large number of pixels satisfying $n = O(\log m)$, Alg.~\ref{algo:block_reg} is consistent.
\end{theorem}
\begin{IEEEproof}
From Lem.~\ref{lemma:MM_exp_consist}, we know MM is exponentially consistent. In particular, for registering $k$ images with multiinformation $\gamma_k$, there exists a universal constant $c$ such that
\[
P_e(\Phi^{k}) \leq 2|\Pi|^k \exp\pth{-n \frac{c}{r^k} \gamma_k^4}.
\]

Using the union bound and Lem.~\ref{lemma:MM_exp_consist}, the error probability of registration using Alg.~\ref{algo:block_reg} is
\begin{equation} \label{eqn:block_regn_union}
P_e(\Phi) \leq 2\frac{m-1}{k-1} |\Pi|^k \exp\pth{-n \frac{c}{r^k} \gamma_k^4}.
\end{equation}
For any fixed $k$, we note that the sufficient condition for consistency is $n = O(\log m)$.
\end{IEEEproof}

From \eqref{eqn:block_regn_union}, we observe that a viable choice of $k$ can be obtained by minimizing the upper bound. Since $\gamma_k$ is proportional to the minimum multiinformation of $k$ images, and since we have no additional knowledge regarding the scaling of such information, one choice is to use the scaling of its trivial upper bound given by $\gamma_k \leq k \log r$ and minimize the resulting expression as in Alg.~\ref{algo:block_reg}.

Thus, we note that using a blockwise MM method, we can perform order-optimal registration of a large number of images with the number of pixels scaling logarithmically with the number of images. This scaling also explains the efficiency of simple pairwise MMI methods in practice.

\subsection{Large-scale Joint Registration and Clustering}

One can directly extend blockwise registration to incorporate clustering as well. In particular, we partition the images into subsets of a fixed size $k$ and depending on the context of clustering, we perform the appropriate joint registration and clustering algorithm defined in Sec.~\ref{sec:joint_cluster_regn}. Then, representative images of each cluster in each subset are collected and the process is repeated to merge clusters across the subsets. Direct analysis similar to Thm.~\ref{thm:necessary_cost} and \ref{thm:sufficient_cost} shows that consistent clustering and registration requires $n = \Theta(\log m)$, similar to universal clustering cost in crowdsourcing \cite{RamanV2017a}. 

The reduction to blockwise registration and clustering also implies that the exhaustive searches are limited. For any $k = O(1)$, the exhaustive search for registration in each subset costs $O(n^{\alpha k}) = O\pth{(\log m)^{\alpha k}}$ computations. Similarly, clustering involves searching over $O(\exp (c k)) = O\pth{(\log m)^c}$ partitions for some constant $c$. Aggregation of subsets requires $O(m)$ block computations. Thus the overall computational cost scales as $O(m (\log m)^{\beta})$, for some $\beta > 0$, i.e.\ polynomially with number of images. However this can still be computationally quite expensive in practice and needs appropriate heuristic adaptations.

\section{Conclusion}

We explored joint clustering and registration of images in a universal setting. Using an information-theoretic analysis of the MMI method for registering two images, we proved asymptotic optimality. We then showed suboptimality of the method in multi-image registration and defined the MM method for registration. We again proved asymptotic optimality using a type counting argument. 

We also defined novel multivariate information functionals to perform consistent, joint clustering and registration of images. For large-scale clustering and registration, we used a blockwise extension to perform consistent clustering and registration. Further, we characterized the sample complexity in terms of the number of pixels per image with respect to the total number of images, and showed order-optimality.

The information-theoretic framework established here to study the performance of the algorithm may be extended to more generic image models that incorporate inter-pixel correlations. Specific knowledge of image transformations could help improve algorithm performance and heuristic adaptations can further reduce computational complexity.

\section*{Acknowledgment}
We thank V.~Misra for discussions and \cite[pp.~xviii--xxi]{Misra2015}.

\bibliographystyle{IEEEtran}
\bibliography{abrv,conf_abrv,lrv_lib}

\end{document}